\documentclass[12pt]{article}
\usepackage{}
\usepackage{geometry}             
\usepackage{graphicx}
\usepackage{amssymb}
\usepackage{amsmath}
\usepackage{epstopdf}
\usepackage{comment}
\usepackage{cite}
\usepackage{caption}
\usepackage{arcs}
\usepackage{newfloat}

\usepackage[hyperindex=true,
          pdfstartview=FitH,
          bookmarksnumbered=true,
          bookmarksopen=true,
          citecolor=blue,
          linkcolor=blue,
          colorlinks=true,
          pdfborder=001,
          unicode]{hyperref}
          
\usepackage[section]{placeins}

\DeclareFloatingEnvironment[placement=htbp,
							within=section]{Quicksheet}

\usepackage[backgroundcolor = shadecolor, linecolor=white]{mdframed}
\usepackage[listings,theorems,skins,breakable]{tcolorbox}
\tcbuselibrary{hooks}

\newtcolorbox{mybox}[2][]{enhanced,
before skip=2mm,after skip=2mm, colback=black!5,colframe=black!50,
attach boxed title to top left={xshift=1cm,yshift*=1mm-\tcboxedtitleheight}, 
boxed title style={frame code={
            \path[fill=tcbcol@back!30!black]
              ([yshift=-1mm,xshift=-1mm]frame.north west)
                arc[start angle=0,end angle=180,radius=1mm]
              ([yshift=-1mm,xshift=1mm]frame.north east)
                arc[start angle=180,end angle=0,radius=1mm];
            \path[left color=tcbcol@back!60!black,right color=tcbcol@back!60!black,
              middle color=tcbcol@back!80!black]
              ([xshift=-2mm]frame.north west) -- ([xshift=2mm]frame.north east)
              [rounded corners=1mm]-- ([xshift=1mm,yshift=-1mm]frame.north east)
              -- (frame.south east) -- (frame.south west)
              -- ([xshift=-1mm,yshift=-1mm]frame.north west)
              [sharp corners]-- cycle;
            },interior engine=empty,
          },
          fonttitle=\bfseries,
          title={#2},#1}

\allowdisplaybreaks

\DeclareMathOperator{\e}{e}
\newcommand{\rd}{{\rm d}}

\renewcommand\({\left(}
\renewcommand\){\right)}

\newcommand\la{\left\langle}
\newcommand\ra{\right\rangle}
\newcommand{\pfrac}[2]{\left(\frac{\partial #1}{\partial #2}\right)}
\newcommand{\bfrac}[2]{\left(\frac{#1}{#2}\right)}

\newcommand{\ppfrac}[3]{\left(\frac{\partial^2 #1}{\partial #2 \partial #3}\right)}
\newcommand{\pfn}[3]{\left(\frac{\partial^{#3} #1}{\partial #2^{#3}}\right)}

\newcommand{\n}{\mathcal{N}}
\newcommand{\A}{\mathcal{A}}
\newcommand{\p}{\mathcal{P}}
\newcommand{\s}{\mathcal{S}}
\newcommand{\q}{\mathcal{Q}}
\renewcommand{\e}{\mathcal{E}}
\newcommand{\en}{\mathfrak{e}}
\newcommand{\sn}{\mathfrak{s}}
\newcommand{\qn}{\mathfrak{q}}
\newcommand{\cn}{\mathfrak{c}}
\newcommand{\gn}{\mathfrak{g}}

\newcommand{\pt}{\partial}

\newcommand{\bluetext}[1]{\textcolor{blue}{#1}}
\newcommand{\purpletext}[1]{\textcolor{purple}{#1}}
\newcommand{\blue}[1]{\textcolor{blue}{#1}}
\renewcommand{\blue}[1]{\textcolor{black}{#1}}  
\newcommand{\cyan}[1]{\textcolor{cyan}{#1}}
\renewcommand{\cyan}[1]{\textcolor{black}{#1}}  
\newcommand{\redtext}[1]{\textcolor{red}{#1}}

\begin{document}
\title{Topological black holes in Einstein-Maxwell and \\
4D conformal gravities revisited}
 
\author{Tao Wang\thanks{
{\em email}: \href{mailto:taowang@mail.nankai.edu.cn}{taowang@mail.nankai.edu.cn}},~
Zhiqiang Zhang\thanks{{\em email}: \href{mailto:2120210176@mail.nankai.edu.cn}
{2120210176@mail.nankai.edu.cn}},~
Xiangqing Kong\thanks{
{\em email}: \href{mailto:2120200165@mail.nankai.edu.cn}{2120200165@mail.nankai.edu.cn}}~
and Liu Zhao\thanks{Correspondence author, {\em email}: 
\href{mailto:lzhao@nankai.edu.cn}{lzhao@nankai.edu.cn}}\\
School of Physics, Nankai University, Tianjin 300071, China}

\date{}
\maketitle
\begin{abstract}
The thermodynamics of charged topological black holes (TBHs) with different 
horizon geometries in $d$-dimensional Einstein-Maxwell and 4-dimensional 
conformal gravities is revisited using the restricted phase space formalism. 
The concept of subsystems for black holes 
is introduced, which enables a precise description for the thermodynamic behaviors 
of (non-compact) black holes with infinitely large horizon area. The concrete 
behaviors can be different for TBHs in the same underlying 
theory but with different horizon geometries, or for those 
with the same horizon geometry but from different underlying theories. 
The high and low temperature limits of the TBHs are 
also considered, and the high temperature limits behave universally as  
low temperature phonon gases. Finally, using the concept of subsystems, some 
conceptual issues in the description for thermal fluctuations in black hole systems 
are clarified, and the relative thermal fluctuations for finite 
subsystems are also analyzed in some detail.

\end{abstract}

\newpage
\tableofcontents
\newpage

\section{Introduction}

Black holes (BHs) in anti-de Sitter (AdS) spacetimes can have event horizons with 
different topologies, which are known as TBHs \cite{lemos1995three,lemos1995two,
lemos1996rotating,aaminneborg1996making,pecca2000thermodynamics,
mann1997pair,SmithMann1997, brill1997thermodynamics, 
mann1997topological, birmingham1999topological,CaiRG}. 
The existence of TBHs with planar, cylindrical and toroidal horizons 
in pure Einstein and Einstein-Maxwell theories was discussed 
in \cite{lemos1995three,lemos1995two,lemos1996rotating}, and the thermodynamics for 
the toroidal AdS black hole is discussed in \cite{pecca2000thermodynamics}. 
More generally, the event horizons 
of TBHs in AdS spacetime can be either compact or non-compact, or even be high \cyan{genus} 
Riemann surfaces \cite{aaminneborg1996making,brill1997thermodynamics,mann1997topological}. 
For the simplest \cyan{genus} zero cases, the local horizon geometries can be either 
maximally symmetric or not. The TBHs which we deal with in the present work are those 
which have maximally symmetric event horizons with 
normalized scalar curvatures $\epsilon=1, 0, -1$, respectively, 
which in turn correspond to spherical, planar or hyperbolic cases. 
Such TBHs in 4-dimensional Einstein-Maxwell theory \cyan{were} found in 
\cite{mann1997pair,brill1997thermodynamics}, and 
their higher dimensional counterparts were presented in \cite{CaiRG}. 
It appears that the existence of TBHs in 
AdS spacetimes is a generic phenomenon, irrespective of the underlying gravity models
and the types of source fields. For instance, TBHs \cyan{exist} in pure Einstein 
gravity \cite{lemos1995three,lemos1995two,aaminneborg1996making,
SmithMann1997,birmingham1999topological}, 
Einstein-Maxwell gravity \cite{lemos1996rotating,mann1997pair,brill1997thermodynamics,
mann1997topological,CaiRG}, Einstein-Maxwell-dilaton 
theory \cite{Mahapatra, Pri}, Einstein-Maxwell-Gauss-Bonnet and Lovelock theories 
\cite{CaiGB,CaiRG}, 
conformal gravity with a Maxwell source \cite{li2013fermi}, etc. 
\cyan{Though one can compactify the transverse spaces for the planar and hyperbolic 
cases in order to keep a finite size 
\cite{brill1997thermodynamics,mann1997topological,nag1988complex}, 
a non-compactified TBH has an infinity horizon ``area'' such that the symbol 
$\omega_{d-2,\epsilon}$ representing the horizon area seems 
only to make sense formally. In this work, we will deal with the 
non-compactified cases when $\epsilon=0,-1$ and try to make perfect sense of the 
thermodynamics of such non-compact TBHs. Of course, the compact TBHs with 
$\epsilon=1$ will also be considered altogether in order to unify the approach
and make comparison between the cases with different choices of $\epsilon$.}

In practice, the compact TBHs have attracted much more interests than the 
non-compact ones, partly because the non-compact nature makes it harder to 
understand the physical properties of the corresponding TBHs. 
For instance, the Bekenstein-Hawking entropy formula points out that the entropy of 
BHs in Einstein gravity has an area law. For non-compact TBHs, 
this amounts to an infinite entropy. \cyan{A potential solution to avoid the infinities 
is to adopt various densities which involves finite factors like 
$M/\omega_{d-2,\epsilon}$ and/or $Q/\omega_{d-2,\epsilon}$ where $M$ and $Q$ represent
respectively the (infinite) mass and charge for the non-compact TBHs. 
However, there is a prerequisite for the density variables to make sense
in thermodynamics, i.e. the Euler homogeneity must hold. Without Euler homogeneity,
the intensive variables would become scale dependent, rendering the density 
variables ill-defined. On the other hand, the whole logic for understanding 
standard thermodynamic systems is based on the existence of finite subsystems. 
In particular, the definition of thermodynamic equilibrium is based on 
the concept of subsystem. In the study of black hole thermodynamics, however, 
the concept of subsystem has not been addressed and put in a proper position 
before. The restricted phase space (RPS) formalism \cite{gao2021restricted, 
wang2021black,gao2022thermodynamics,zhao2022thermodynamics,kong2022restricted} 
places Euler homogeneity in a position of utmost importance,} which enables us 
to study the local thermodynamics of the non-compact TBHs. \cyan{This formalism also 
allows for an easy introduction of finite subsystems for the non-compact TBHs.} 
A detailed discussion on the importance of Euler homogeneity and the concept 
of subsystems will be provided in Section 2.

In this work, we will study the local RPS thermodynamics of the TBHs in 
$d$-dimensional Einstein-Maxwell theory and in 4-dimensional conformal gravity 
with a Maxwell source, with emphasis paid toward the comparison between different 
horizon geometries. Our results indicate that, for TBHs in the 
same underlying theory, the local thermodynamic behaviors are quite different 
for TBHs in the same theory but with different horizon geometries. On the other hand, 
in the high temperature limit, all TBHs in $d$-dimensional AdS spacetime behave 
exactly like the quantum phonon gas in nonmetallic crystals residing in 
$(d-2)$-dimensional flat space -- a feature first discovered for spherical 
Tangherlini-AdS black holes\cite{kong2022high} and subsequently generalized to 
the case of charged spherically symmetric AdS black hole
in conformal gravity\cite{kong2022restrictedconformal}. Now the same feature is 
further validated for charged Tangherlini-AdS TBHs in Einstein-Maxwell theory 
in any dimensions $d\geq4$ and for charged AdS TBHs in conformal gravity in 4 dimensions, 
regardless of the \cyan{horizon} geometries. 
It is natural to make the conjecture that the above AdS/phonon gas correspondence 
may be a universal feature for BHs in AdS spacetime of any dimension $d\geq 4$, 
regardless of the horizon geometry and the underlying model of gravity.  
More detailed behaviors for TBHs in the above two theories will be introduced in 
the main text. 

Throughout this paper, we will be working in units $c=\hbar=k_{\rm B}=1$, 
while leaving the Newton constant $G$ intact, because $G$ needs to be variable 
in the RPS formalism. The static black hole solutions will be presented using 
Schwarzschild like coordinates.

This paper is organized as follows. Section 2 provides a discussion on the concept of subsystems and compares the differences between existing black hole thermodynamic formalisms. Section 3 is devoted to a thorough study of 
TBHs in Einstein-Maxwell theory. The introduction of subsystems, the analogues of Clapeyron 
and Ehrenfest equations and the high and low temperature limits of the TBHs are among the 
major subjects of concern, besides the very detailed study of thermodynamic processes 
with emphasis on the comparison between the effects of different horizon geometries.
Section 4 presents the parallel study for TBHs in $4d$ conformal gravity in much 
more brevity. In Section 5, we discuss the local thermodynamic fluctuations 
in small closed subsystems of the black holes and 
clarify some of the conceptual issues in existing treatments of black hole fluctuations 
using either Smoluchowski formula or the thermodynamic geometric approaches. 
Finally, in Section 6, we present the summary of the 
results and make some discussions.

\section{Formalisms and subsystems of black hole thermodynamics}

An outstanding feature of thermodynamic systems is the existence of subsystems
with essentially the same macroscopic behavior, albeit with different 
microscopic degrees of freedom. The concept of thermodynamic equilibrium for ordinary
thermodynamic systems is defined by the system-wide homogeneity of 
the intensive thermodynamic variables such as the temperature $T$, the pressure $P$ and 
the chemical potential $\mu$, and the equilibrium conditions are described as
\[
T_A=T_B,\quad P_A=P_B,\quad \mu_A=\mu_B
\]
for arbitrarily chosen subsystems $A$ and $B$. If we consider charged BHs as
thermodynamic systems, there should be an additional equilibrium condition 
which describes the system-wide homogeneity of the electric potential,
\[
\hat\Phi_A=\hat\Phi_B,
\]
and the condition over the pressure may or may not be present, depending on the 
choice of formalisms for black hole thermodynamics.  Clearly, we need to identify 
what the subsystems are for BHs in order to make sense of the above 
equilibrium conditions.

In thermodynamic descriptions for ordinary matters, defining a subsystem has never 
been a problem. The whole logic of the classical thermodynamics is established on 
the recognition of the following fact, i.e. the thermodynamic variables can be arranged 
in two groups, with one group consists of variables which are uniform and take the same 
values on different subsystems, and the other group consists of variables which 
are dependent on the matter content within each subsystem 
(i.e. proportional to the number of particles in each subsystem) and are additive when 
different subsystems are combined together. The Euler homogeneity is a prerequisite 
for classifying thermodynamic variables into the uniform and additive groups. 

One may wonder why an infinite area of the event horizon could lead to problems
in describing the thermodynamic properties. For ordinary macroscopic systems in 
thermodynamic limit, the particle number and hence the internal energy as well 
as the entropy also diverge. But this has never constituted a difficulty in understanding
thermodynamic properties. The key point lies in that, one can always take a finite 
subsystem and study the thermodynamic behaviors thereof. Thanks to the Euler homogeneity and 
the extensivity of ordinary thermodynamics, any subsystem behaves the same 
\cite{callen1998thermodynamics}. 
In the case of non-compact TBHs, the major obstacle comes from the lack of extensivity
in the usual thermodynamic description in either the traditional 
formalism\cite{bekenstein1972black, bekenstein1973black, bardeen1973four, 
smarr1973mass, hawking1975particle, bekenstein1975statistical} or the 
extended phase space (EPS) formalism \cite{kastor2009enthalpy, dolan2010cosmological, 
dolan2011pressure, dolan2011compressibility, kubizvnak2012p, gunasekaran2012extended, 
belhaj2012thermodynamics, hendi2013extended, chen2013p,zhao2013critical, poshteh2013phase, 
altamirano2013reentrant, cai2013pv, belhaj2013thermodynamical, 
xu2014gauss, zou2014critical, 
altamirano2014thermodynamics, wei2014triple, kubizvnak2015black, zou2014critical-gb, 
xu2014extended, zhang2015phase, kubizvnak2017black, lemos2018black} (including 
the holographic variant thereof \cite{cong2021thermodynamics, visser2022holographic,
cong2022holographic}). 
Some authors even think of the lack of extensivity as one of the distinguished \cyan{features} 
of black hole thermodynamics \cite{1910.03123,2208.04473,2211.05989}. 
The real problem lies in that, in the absence of Euler homogeneity, 
the thermodynamic behavior becomes scale dependent, putting two subsystems together 
yields a larger system with different thermodynamic behavior. Therefore, studying subsystems 
does not resolve the problem for infinitely large systems in the lack of 
Euler homogeneity. 

However, the Euler homogeneity has rarely been discussed seriously in the studies of 
black hole thermodynamics before the restricted phase space (RPS) 
formalism was proposed recently in \cite{gao2021restricted, 
wang2021black,gao2022thermodynamics,zhao2022thermodynamics,kong2022restricted,kong2022high}. 
Some exceptional works, e.g. \cite{Tian1,Tian2} and \cite{visser2022holographic}, 
did make some discussions on Euler homogeneity, but these works depend heavily on the 
holographic interpretation which constrains the range of applicability. Moreover, 
the parameter describing the microscopic degrees of freedom for the BHs is absent
in the works \cite{Tian1,Tian2}, and the Euler homogeneity 
in \cite{visser2022holographic} is incomplete (there is a pair of thermodynamic 
variables $(\mathcal{P,V})$ but these variables did not appear in the 
so-called ``Euler relation'').
The RPS formalism is distinguished from that of \cite{visser2022holographic} in that
the cosmological constant is not taken as a variable, thus the variables $(\mathcal{P,V})$
are absent. Moreover, the new variables $N,\mu$ are defined {\em independently}
and are interpreted as the number of black hole molecules and the conjugate 
chemical potential,
\begin{align}\label{Nmu}
N=\dfrac{L^{d-2} \omega_{d-2,\epsilon}}{G},\quad\quad
\mu=\dfrac{GTI_{E}}{L^{d-2} \omega_{d-2,\epsilon}},
\end{align} 
where $L$ is a finite length scale which is chosen to contain all microscopic 
degrees of freedom for the BHs\cite{wang2021black}, whose value remains to be 
arbitrary besides the above requirement, and $I_E$ is the on-shell 
Euclidean action \cite{gibbons1977action,chamblin1999charged}. 
The factors $\omega_{d-2,\epsilon}$ in $N$ and $\mu$ were absorbed in $L$ in our previous 
works on the RPS formalism for compact BHs. However, for non-compact TBHs, these 
factors need to be made explicit in order to keep $\mu$ finite. 

For AdS BHs, $N$ is proportional to the central charge of the dual CFT in the framework of 
AdS/CFT correspondence \cite{visser2022holographic,gao2021restricted,
wang2021black}, but its best interpretation may be due to 't Hooft \cite{Hooft}, 
who suggested that any piece of roughly the size of Planck area (which equals  $1/G$) 
on the event horizon corresponds to a miscroscopic degree of freedom of the black hole. 
Since we do not know what exactly a black hole molecule is, there is a freedom 
in choosing the coefficient of proportionality $L$ between $N$ and $1/G$. This situation 
is quite similar to the study of thermodynamics of ordinary matter systems, 
in which the precise nature of individual molecules does not matter, and 
the total number of molecules can be taken arbitrarily as long as the whole system 
remains macroscopic, which means that $L$ should be sufficiently large and remains constant in our present case.

\begin{Quicksheet}
\begin{tcolorbox}[colback=red!5!white,colframe=red!65!black,
lower separated=true,
lefthand ratio=0.5] 
{\bf Traditional formalism:}  \purpletext{$\Lambda, G$ are both constant}

\bluetext{Bekenstein, Bardeen, Carter, Hawking {\em et al}} 
\begin{itemize}
\item Range of applicability: any black hole spacetime;
\item Role of mass: internal energy ($E=M$);
\item The first law: 
\[
\rd M = T\rd S + \Phi\rd Q;
\]
\item Mass formula: the Smarr relation 
\[
M=\dfrac{d-2}{d-3}TS + \Phi Q +\frac{\Lambda \Phi S^2}{\pi^2(d-1)Q};
\]
\item Phase transitions: not touched upon.
\end{itemize}
\end{tcolorbox}
\end{Quicksheet}

To make comparison between different formalisms, 
we created several quick sheets, listing the main features of some of the 
major formalisms of black hole thermodynamics, including the range of applicability,
the interpretation of the mass, and the form of the first law and Smarr/Euler relations 
in each formalism. While writing down the first laws and the mass formulae, 
we have taken the $d$-dimensional spherically symmetric Tangherlini-RN-AdS 
black hole as an example case.
It can be seen that only the traditional and the RPS formalisms 
are applicable to all black hole spacetimes irrespective of the asymptotic behaviors. 
Moreover, at fixed Newton constant $G$, the RPS formalism falls back to the 
traditional formalism, except that the Euler relation still needs the contribution from 
the product of $N$ and $\mu$. This situation is fully consistent with the ordinary 
thermodynamics for closed systems.

\begin{Quicksheet}
\begin{tcolorbox}[colback=red!5!white,colframe=red!65!black,
lower separated=true,
lefthand ratio=0.5] 
{\bf EPS formalism:} \purpletext{$\Lambda$ is variable but $G$ is constant}

\bluetext{Kastor, Ray, Traschen {\em et al}} 
\begin{itemize}
\item Range of applicability:  AdS black holes;
\item Role of mass: enthalpy ($H=M$);
\item The first law:  ({$P=-\Lambda/8\pi$})
\[
\rd M = T\rd S + \Phi\rd Q + \bluetext{V\rd P};
\]
\item Mass formula: the generalized Smarr relation
\[
M=\dfrac{d-2}{d-3}TS+\Phi Q-\dfrac{2}{d-3}PV;
\]
\item Phase transitions: $P-v$ criticalities studied extensively, 
but not working for non-spherical BHs. 
\end{itemize}
\end{tcolorbox}
\end{Quicksheet}

\begin{Quicksheet}[ht]
\begin{tcolorbox}[colback=red!5!white,colframe=red!65!black,
lower separated=true,
lefthand ratio=0.5] 
{\bf Holographic EPS formalism:} \purpletext{$\Lambda, G$ are both variable}

\bluetext{Visser {\it et al}} 
\begin{itemize}
\item Range of applicability:  AdS black holes;
\item Role of mass: internal energy ($E=M$);
\item The first law: 
\[
\rd M = T\rd S + \hat\Phi\hat\rd Q 
\redtext{-\p\rd\mathcal{V}}+\bluetext{\mu\rd C};
\]
\item Mass formula: Smarr relation disguised as Euler relation 
\[
M=TS  + \hat\Phi \hat Q +\mu C;
\]
\item Extra variables: 
{$C = \ell^{d-2}/G$} is the central charge of the dual CFT, and
$\mu$ is defined using the mass formula,
\[
{\mu \equiv (M-TS- \hat\Phi \hat Q)/C}.
\] 
\item Phase transitions: $\mu-C$ criticalities
\end{itemize}
\end{tcolorbox}
\end{Quicksheet}

\begin{Quicksheet}[ht]
\begin{tcolorbox}[colback=red!5!white,colframe=red!65!black,
lower separated=true,
lefthand ratio=0.5] 
{\bf RPS formalism:} \purpletext{$\Lambda$ is constant but $G$ is variable}

{\bluetext{Our group}} 
\begin{itemize}
\item Range of applicability: any black hole spacetime;
\item Role of mass: internal energy ($E=M$);
\item The first law: 
\[
\rd M = T\rd S + \hat\Phi\hat\rd Q  + \blue{\mu\rd N};
\]
\item Mass formula: the true Euler relation
\[
M= TS+\hat\Phi\hat Q +\mu N;
\]
\item Extra variables: defined independent of holographic 
duality and mass formula,
\[
{N=\frac{L^{d-2} \omega_{d-2,\epsilon}}{G},\quad 
\mu=\frac{GTI_{E}}{L^{d-2} \omega_{d-2,\epsilon}}}.
\] 
\item Phase transitions: $T-S$~transition for AdS BHs as well as 
Hawking-Page like transitions.
\end{itemize}
\end{tcolorbox}
\end{Quicksheet}

It should mentioned that, the $\mu-C$ criticality described in e.g. 
\cite{cong2022holographic} under the holographic EPS formalism contains some 
conceptual issues. Fig.7 of \cite{cong2022holographic} depicted 
the central charge $C$ as a function in the chemical potential $\mu$, and the authors 
explicitly mentioned that there is an equal area law on such curves. This 
description is incorrect, because the areas in the equal area law have 
physical interpretations. For $P-V$ criticalities in ordinary thermodynamic systems, 
the areas correspond to works done on the system following two different 
isothermal processes, and for $T-S$ criticalities described in \cite{gao2021restricted, 
wang2021black,gao2022thermodynamics}, the areas correspond to heats absorbed by the 
BHs in two different isocharge processes. The areas depicted in Fig.7 of 
\cite{cong2022holographic} are related to $\displaystyle \int C\rd\mu$, 
whose meaning looks obscure. On the other hand, $\mu$ should be regarded as an 
intensive macroscopic state function in $(T,\hat{\Phi})$, thanks to the 
Gibbs-Duhem relation. When $\mu$ is multiply-valued, only the lowest values 
correspond to states in a thermodynamically stable phase.

Since the RPS formalism contains a variable Newton constant $G$, it looks 
necessary to interpret our take on this variable. Essentially, BHs as thermal objects
must be quantum. As such, it should not be surprising that the gravitational 
coupling constant could be running with the energy scale along the renormalization group
orbit. On the other hand, we have no objection to the current observational picture 
of our universe with a fixed Newton constant. This may be interpreted as a infrared 
fixed point for $G$, and, in a universe with fixed $G$, the number of black hole molecules
$N$ is necessarily fixed also, hence any BH in such a universe must be considered 
as closed thermodynamic system. Anyway, this will not affect our discussions on the 
phase structures, because, when studying phase transitions, the number of molecules
in the thermodynamic system under investigation should always be kept fixed.

Once the complete Euler homogeneity is established using the RPS formalism, 
we are free to take some finite parts on the event horizon of the TBHs as 
subsystems. The local thermodynamic description for the subsystems 
is free of the ambiguities brought about by the 
infinite area. Following this route, we will be able to make 
comparison between the cases with different horizon geometries.

For convenience and consistency of later discussions, 
we provide some notational conventions about subsystems.
The subsystems which we are interested in are finite subsystems containing 
a finite number $\n$ of black hole molecules and have finite entropy $\s$ and 
electric charge $\q$. According to eq.\eqref{Nmu}, a finite number of black hole 
molecules cannot be achieved for non-compact TBHs by choosing an alternative value 
for $L$ and/or $G$, provided they are still kept finite. The only way to take a 
finite subsystem from the TBHs is to replace $\omega_{d-2,\epsilon}$ with 
the area $\A$ of a finite part of \cyan{the $(d-2)$-dimensional transverse submanifold} 
$\Sigma_{d-2,\epsilon}$  
(let us stress that, just like $\omega_{d-2,\epsilon}$, $\A$ is also a dimensionless 
quantity). This implies that the subsystem is actually taken to be a finite piece 
on the event horizon of area $\A r_0^{d-2}$, where $r_0$ represents the 
radius of the horizon. This is in agreement with 't Hooft's picture for black hole 
micro states \cite{Hooft}, which states that the micro states of BHs can be 
regarded to be located on the event horizons.

The number of black hole molecules contained in the above subsystem reads 
$\n= L^{d-2}\A/G$, which need to obey the condition $1\ll\n\ll N$, and the entropy of the 
subsystem is $\s=\A r_0^{d-2}/4G$. Likewise, 
the internal energy $\e$ and the electric charge $\q$ of the subsystem must also 
be proportional to $\A$. It is extremely important 
to realize that, due to Euler homogeneity, the behaviors of any meaningful 
subsystems are the same for a system in thermodynamic equilibrium, provided only 
the quantities which are zeroth order homogeneous functions in the additive variables 
are concerned.

\section{TBHs in Einstein-Maxwell theory}

\subsection{The metric and the black hole observables}

The action of Einstein-Maxwell theory in $d$-dimensions reads
\begin{align}
I=\dfrac{1}{16\pi G}\int\rd^{d}x\sqrt{-g}(R-2\Lambda)
-\dfrac{1}{16\pi}\int\rd^{d}x\sqrt{-g}F_{\mu\nu}F^{\mu\nu},
\end{align}
where the negative cosmological constant $\Lambda$ is related to the AdS radius $\ell$ 
via $\Lambda =-(d-1)(d-2)/2\ell^{2}$. Unlike in the EPS formalism, which takes the 
AdS radius $\ell$ as one of the thermodynamic variables, the RPS formalism 
keeps $\ell$ as a fixed constant. This makes it possible to 
establish complete Euler homogeneity which is missing in 
alternative considerations.
 
The line element and the Maxwell field $A_\mu$ presented in \cite{CaiRG} 
are given as follows,
\begin{align}
&\rd s^{2}=-f(r)\rd t^{2}+\dfrac{\rd r^{2}}{f(r)}+r^{2}\rd\Omega_{d-2,\epsilon}^{2},\\
&A_{\mu}\rd x^{\mu}=\Phi(r)\rd t,
\end{align}
where $\epsilon=1,0,-1$, $\rd\Omega_{d-2,\epsilon}^{2}=\gamma^{(\epsilon)}_{ab}\rd\theta^a
\rd\theta^b$ represents the line element of the 
$(d-2)$-dimensional submanifold $\Sigma_{d-2,\epsilon}$ with normalized 
scalar curvature $\epsilon$ and ``area'' (or, more precisely, solid angle because 
of its dimensionless nature) 
\[
\omega_{d-2,\epsilon}=\int \rd^{d-2}\theta \sqrt{{\rm det}(\gamma^{(\epsilon)}_{ab})},
\]
and 
\begin{align}
f(r)&=\epsilon-\left(\dfrac{r_{g}}{r}\right)^{d-3}+\left(\dfrac{r_{Q}}{r}\right)^{2(d-3)}
+\dfrac{r^{2}}{\ell^{2}},\\
\Phi(r)&=\dfrac{4\pi\,Q}{(d-3)\omega_{d-2,\epsilon}\,r^{d-3}},
\end{align}
in which 
\begin{align}
r_{g}=\left[\dfrac{16\pi GM}{(d-2)\omega_{d-2,\epsilon}}\right]^{\frac{1}{d-3}}, \quad\quad
r_{Q}=\left[\dfrac{32\pi^{2}GQ^{2}}{(d-2)(d-3)
\left(\omega_{d-2,\epsilon}\right)^{2}}\right]^{\frac{1}{2(d-3)}}.
\end{align}

The black holes encoded in the above line element is known as the Tangherlini-RN-AdS TBHs. 
For $\epsilon=1$, $\omega_{d-2,\epsilon}$ corresponds to the area of the unit 
$(d-2)$-sphere, which is finite and makes perfect sense. 
While for $\epsilon=0,-1$, respectively, $\omega_{d-2,\epsilon}$ is 
in fact divergent. However, it is assumed that the ratios $M/\omega_{d-2,\epsilon}$ and
$Q/\omega_{d-2,\epsilon}$ are both finite, so that the metric and the electric potential 
for the TBHs still make sense. Since there is a pole at $d=3$ in the expression 
for $r_Q$, all we can do is to consider the spacetimes with $d\geq 4$. 

Let $r_0$ be a real positive root of $f(r)$ which corresponds to the position of
the event horizon. In the RPS formalism, the internal energy $E$ of the BHs 
is identified with the mass $M$, which can be expressed as an explicit function in 
$r_0, Q, G$ by solving the equation $f(r_0)=0$,  
\begin{align}
E&=M=\dfrac{(d-2)\omega_{d-2,\epsilon}\,r_{0}^{d-3}}{16\,\pi\,G}
\left(\epsilon+\dfrac{r_{0}^{2}}{\ell^{2}}\right)
+\dfrac{2\,\pi\,Q^{2}}
{(d-3)\omega_{d-2,\epsilon} r_{0}^{d-3}}.
\end{align}
The Bekenstein-Hawking entropy and the Hawking temperature corresponding to
the above solution read
\begin{align}
S&=\dfrac{\omega_{d-2,\epsilon}r_{0}^{d-2}}{4\,G}, \qquad 
T=\frac{1}{4 \pi r_0} \left[(d-3) \epsilon -\frac{32 \pi^2 G Q^2 }
{(d-2) \(\omega_{d-2,\epsilon}\)^{2} r_0^{2(d-3)}}+\frac{(d-1) r_0^2}{\ell^2}\right].
\end{align} 

In the RPS formalism, we need to introduce a new pair of thermodynamic quantities
$(N,\mu)$ as given in eq.\eqref{Nmu}, in which the on-shell 
Euclidean action $I_E$ is evaluated to be \cite{gibbons1977action,chamblin1999charged}
\begin{align}
I_E=-\frac{\omega_{d-2,\epsilon} r_0^{d} 
\left[(d-3) (d-2) \left(\omega_{d-2,\epsilon}\right)^2 r_0^{2 d}
\left(1-\epsilon \frac{\ell^2}{r_0^2}\right)
+32 \pi ^2 G \ell^2 Q^2 r_0^4 \right]}
{4 (d-3) G \left[(d-2) \left(\omega_{d-2,\epsilon}\right)^2 r_0^{2 d}
\left((d-3) \epsilon \ell^2+(d-1) r_0^2\right)
-32 \pi ^2 G \ell^2 Q^2 r_0^6\right]}.
\end{align}
Although the expression for $I_E$ looks quite complicated (and we do not use
the above complicated expression for $I_E$ in the rest of this work), 
the corresponding expression for the chemical potential is much simpler,
\begin{align}
\mu&=\frac{r_0^{d-3}}{16 \pi L^{d-2}}\epsilon
-\frac{2\pi G}{(d-2)(d-3)\left(\omega_{d-2,\epsilon}\right)^2 L^{d-2}r_0^{d-3} }Q^2
-\frac{r_0^{d-1}}{16 \pi L^{d-2} \ell^2}. \label{muorig}
\end{align}

In the RPS formalism, the electric charge and potential need to be rescaled, 
so that the Newton constant becomes an overall factor in front of the full action. 
The reason for this rescaling has been explained in \cite{gao2021restricted}. 
The rescaled electric charge and potential are given as
\begin{align}
\hat Q=\dfrac{Q\,L^{(d-2)/2}}{\sqrt{G}}, \qquad 
\hat\Phi=\dfrac{4\pi\,Q\,\sqrt{G}}{(d-3)\omega_{d-2,\epsilon}\,r^{d-3}\,L^{(d-2)/2}}.
\end{align}

It should be reminded that, due to the presence of the formal parameter 
$\omega_{d-2,\epsilon}$, the thermodynamic quantities introduced above make perfect sense 
only for compact BHs. Nevertheless, if one keeps the formal notation 
$\omega_{d-2,\epsilon}$ as if it is a finite constant, it can be checked with ease that 
the following basic thermodynamic relations hold for any TBH, be it compact or not,
\begin{align}
&E=T\,S+\hat\Phi\,\hat Q+\mu\,N, \label{Euler} \\
&\rd E=T\,\rd S+\hat\Phi\,\rd \hat Q+\mu\,\rd N,\\
&S\rd T+\hat{Q}\rd \hat{\Phi}+N\rd\mu=0. \label{GDR}
\end{align}
The first one of these relations is exactly the Euler relation, the second one is the 
first law of thermodynamics, while the third relation follows from the first and the 
second, and is known as the Gibbs-Duhem relation, which is very important in
ordinary thermodynamics. These relations signify the 
first order homogeneity of $E$ and zeroth order homogeneity of $T,\hat\Phi,\mu$ 
in terms of $(S,\hat Q, N)$.  Such homogeneity behaviors are absent in the traditional
and the EPS formalisms of black hole thermodynamics, which prevents the analysis of
thermodynamic properties of TBHs from the point of view of subsystems. Now in the RPS
formalism, we are able to do so. 

\subsection{Subsystems and local thermodynamic variables}

The importance of the concept of subsystems in thermodynamics and its basic definition have been explained in Section 2. To standardize the operations, we now introduce the dimensionless 
mean values of the additive variables per black hole molecule 
(with appropriate rescaling) for the subsystem as follows,
\begin{align}
e&\equiv 2\pi\ell a^{d-2}\bfrac{\e}{\n}=2\pi\ell a^{d-2}\bfrac{E}{N }, \label{edef}\\
s&\equiv 4 a^{d-2} \bfrac{\s}{\n}=4a^{d-2} \bfrac{S}{N},
\label{sN}\\
q&\equiv 2\pi\ell \bfrac{a}{\ell}^{(d-2)/2}\bfrac{\q}{\n}
=2\pi\ell \bfrac{a}{\ell}^{(d-2)/2}\bfrac{\hat Q}{N}, \label{qN}
\end{align}
together with the dimensionless uniform variables which are also rescaled 
accordingly,
\begin{align}
\bar{\mu}&\equiv 2\pi\ell a^{d-2}\,\mu,\quad
t\equiv \frac{\pi\ell\,T}{2}, \quad 
\phi\equiv ({\ell}{a})^{(d-2)/2}\hat\Phi. \label{tdef}
\end{align}
In these definitions, we have introduced $a\equiv L/\ell$. 
The numeric coefficients such as $2\pi$, $\pi/2$, $4$ and powers of $a$ 
are not strictly necessary, but are included as a convention, which greatly helps 
in simplifying the thermodynamic functions and equations of states to be given shortly. 
Because of the introduction of the factor $1/\n$ in the definitions of $e, s, q$, 
none of the above variables is additive. By slightly abuse of terminologies, we 
still refer to $e, s, q$ respectively as (the local densities of) the internal energy,
entropy and electric charge, and to $t, \phi,\bar\mu$ simply as the temperature, 
electric potential and chemical potential,  
regardless of the dimensionless natures thereof. 

Now, taking $s,q$ as fundamental independent variables, the other dimensionless 
variables can be written as explicit functions in $(s,q)$, 
\begin{align}
e&=\frac{d-2}{8}s^{\frac{d-3}{d-2}}\epsilon
+\frac{1}{d-3}s^{-\frac{d-3}{d-2}}\,q^{2}
+\frac{d-2}{8}\,s^{\frac{d-1}{d-2}} ,\label{e}
\\
t&=\frac{d-3}{8}s^{-\frac{1}{d-2}}\epsilon
-\frac{1}{d-2}s^{-\frac{2d-5}{d-2}}\,q^{2}
+\frac{d-1}{8}\,s^{\frac{1}{d-2}},\label{t}
\\
\phi&=\dfrac{2}{d-3} \,s^{-\frac{d-3}{d-2}}\,q,\label{phi}\\
\bar{\mu}&=\frac{1}{8} s^{\frac{d-3}{d-2}}\epsilon
-\frac{1}{(d-2)(d-3)}s^{-\frac{d-3}{d-2}}\,q^{2}
-\frac{1}{8}\, s^{\frac{d-1}{d-2}}.\label{barmu}
\end{align}
Eq.\eqref{e} presents the (density of) the thermodynamic potential $e$ with 
adapted independent variables $(s,q)$, and eqs.\eqref{t}-\eqref{barmu} provide 
the corresponding equations of states. Notice that all these equations are 
independent of the parameter $a$, which reflects the fact that the choice of $L$ 
is arbitrary and is irrelevant to the local thermodynamic behaviors. The 
above equations are also independent of the number of black hole molecules $\n$
contained in the subsystem. This is a characteristic feature of ordinary 
thermodynamic systems known as the law of corresponding states, however, it is 
absent in other formalisms for black hole thermodynamics. 

Except the $\phi-q$ relation \eqref{phi}, $q$ always appear in squared form 
in the other thermodynamic functions. Therefore, it suffices to consider only 
the cases with $q\geq 0$, which also ensures $\phi\geq 0$ because $s$ can never be 
negative.

Besides the above independent variables and local thermodynamic functions, 
we also need the dimensionless local density of the Helmholtz free energy $f=e-ts$, 
specifically for describing possible $t-s$ criticalities. We have, explicitly,
\begin{align}
f&=\frac{1}{8}s^{\frac{d-3}{d-2}}\epsilon
+\frac{2d-5}{(d-2)(d-3)}s^{-\frac{d-3}{d-2}}\,q^{2}
-\frac{1}{8}\,s^{\frac{d-1}{d-2}}. \label{f}
\end{align}
Please be reminded that, $f$ should be regarded as an implicit function in $(t,q)$, 
rather than a function in $(s,q)$. Please also take a notice on 
the explicit $\epsilon$-dependences of the expressions for 
$e, t,\bar\mu$ and $f$, which indicates the influence of horizon geometries 
on the thermodynamic functions.

Using the above dimensionless local variables, the Euler relation and the 
Gibbs-Duhem relation can be rewritten as
\begin{align}
&e=ts+\phi q+\bar{\mu},\label{euler}\\
&s\rd t+q\rd \phi+\rd\bar{\mu}=0. \label{GB}
\end{align}
Using these relations we also have
\begin{align}
&f=\phi q+\bar{\mu},\\
&\rd f= - s\, \rd t + \phi\, \rd q. \label{df}
\end{align}
The last equation will be useful for analyzing the critical behaviors in the isocharge
$T-S$ processes in the case $\epsilon=1$. 

\subsection{Bounds over independent parameters}

Charged BHs subject to some bound -- known as the Bogomol'nyi bound -- 
over the mass and charge in order to ensure the 
existence of event horizons. In the context of thermodynamics, the Bogomol'nyi bound 
arises from the requirement of non-negativity for the temperature. It can be seen from
eq.\eqref{t} that, for $d\geq 4$, the coefficients in front of the terms 
involving $\epsilon$ and $q^2$ can both be negative, while the last term in \eqref{t}
is strictly non-negative. Thus the requirement of non-negativity for $t$ imposes real
bound over the independent variables $s$ and $q$, i.e.
\begin{align}
q^{2}\leq \frac{d-2}{8}\left[(d-3)s^{\frac{2(d-3)}{d-2}}\epsilon
+(d-1)s^{2}\right]. \label{bogo1}
\end{align}
When the above bound saturates, the temperature vanishes, which corresponds to 
either a completely evaporated black hole with no leftover ($s=0, q=0$), 
or to an extremal black hole remnant with nonvanishing $s$ and $q$. 
This implies that, for $\epsilon=-1$, the state with $s=0$ is inaccessible 
even asymptotically, while for $\epsilon=0,1$, the state with $s=0$ is accessible 
at least asymptotically. 

We can make a change of variable $q\to\phi$ in \eqref{t} by use of eq.\eqref{phi}. 
Then the Bogomol'nyi bound can also be written as
\begin{align}
\phi^{2} \leq \frac{d-2}{2(d-3)}\left[\epsilon
+\frac{d-1}{d-3}s^{\frac{2}{d-2}}\right].\label{bogo2}
\end{align} 

One might also consider the non-negativity of the mass or internal energy as the 
source of another bound, but this \cyan{is} not the case in AdS spacetimes, because even the 
vacuum has a negative energy density. Therefore, the Bogomol'nyi bound remains to be 
the only physical bound over the range of permitted independent variables.

\subsection{(Non)existence of critical points and HP-like transitions}

A remarkable feature of the RPS description for AdS BHs with compact event horizons 
is the existence of isocharge (or iso-angular-momenta) $T-S$ phase transitions 
at supercritical temperatures \cite{gao2021restricted,gao2022thermodynamics,kong2022restricted}. 
It is natural to ask whether the same feature persists 
in the cases with non-compact horizons. 

The analytical expressions for the local thermodynamic functions presented in Section 3.2 
allow us to study the existence of critical points in the isocharge $t-s$ processes 
for each \cyan{choice} of spacetime dimension and horizon geometry. The critical 
point $(s_c,q_c)$, if exists, must obey the following equations,
\begin{align}\label{criticaleq}
\pfrac{t}{s}_{q}=0,\quad \pfn{t}{s}{2}_{q}=0.
\end{align}
Using eq.\eqref{t}, it can be seen that the above system of equations does not 
have any nonzero solution for $\epsilon=0,-1$, thus excluding the existence 
of $t-s$ criticalities for planar and hyperbolic black holes in any dimensions. 
For spherical black holes, there always exists a critical point in any 
dimension $d\geq4$. The position of the critical point for $4\leq d\leq 10$ 
together with the corresponding critical temperatures are 
listed in Table 1. The existence/nonexistence of critical points 
makes a sharp difference between BHs with compact and non-compact event horizons. 
The difference begins to manifest in the planar case ($\epsilon=0$), 
for which the critical point equations \eqref{criticaleq} have a 
unique solution $(s_c,q_c)=(0,0)$, which should not be regarded as a critical point. 
For the hyperbolic case ($\epsilon=-1$), the critical point equations \eqref{criticaleq}
do not admit any real non-negative solution. 

\begin{table}[htbp]
\centering
\caption{Critical parameters for $\epsilon=1, 4\leq d\leq 10$}
\begin{tabular}{c|ccccccc}
\hline
$d$ & 4 & 5 & 6 & 7 & 8 & 9 & 10 \\[3pt]
\hline 
$s_c$ & 
$\frac{1}{6}$ & 
$\frac{\sqrt{3}}{9}$ &
$\frac{81}{400}$ &
$\frac{128}{255}\sqrt{\frac{2}{15}}$ &
$\frac{15625}{74088}$ &
$\frac{2187\sqrt{14}}{38416}$ &
$\frac{5764801}{26873856}$ \\[6pt]
$q_c$ &
$\frac{1}{12}$ &
$\frac{\sqrt{5}}{30}$ &
$\frac{27}{80}\sqrt{\frac{3}{70}}$ &
$\frac{32\sqrt{2}}{675}$ &
$\frac{3125}{7056}\sqrt{\frac{5}{231}}$ &
$\frac{729}{5488}\sqrt{\frac{3}{13}}$ &
$\frac{823543}{8757952}\sqrt{\frac{7}{15}}$\\[6pt]
$t_c$ &
$\frac{\sqrt{6}}{6}$ &
$\frac{2\sqrt{3}}{5}$ &
$\frac{3\sqrt{5}}{7}$ &
$\frac{2}{3}\sqrt{\frac{10}{3}}$ &
$\frac{5\sqrt{42}}{22}$ &
$\frac{6\sqrt{14}}{13}$ &
$\frac{7\sqrt{2}}{5}$\\
\hline
\end{tabular}
\end{table}

Another remarkable feature of the RPS description for charged AdS BHs with compact 
event horizons is the existence of HP-like transitions signified by the  
zero of the chemical potential \cite{gao2021restricted,gao2022thermodynamics,
kong2022restricted}. We call such transitions HP-like, because the HP transition 
is only defined in the neutral case and is interpreted as the transition between 
the neutral AdS black hole and a pure thermal gas\cite{hawking1983thermodynamics}. 
In the presence of electric charge, the zero for the chemical potential 
still exists, but we simply do not understand what a charged thermal gas is. 

Now, according to eq.\eqref{barmu}, the HP-like transition could arise only for TBHs 
with compact horizons, because only when $\epsilon=1$, $\bar\mu$ could have zero(s)
at real positive $(s,q)$. The zero occurs at $r_0=\ell$ if $q=0$, thanks to 
eq.\eqref{muorig}, the corresponding temperature reads $T_{\rm HP}=\frac{d-2}{2\pi\ell}$,
or $t_{\rm HP}=\frac{d-2}{4}$. Therefore, for compact neutral AdS BHs, the radius of the 
event horizon could not be identical to the AdS radius, otherwise the BHs would 
become pure thermal gases due to the HP transition. 
For $\epsilon=0$, $\bar\mu$ has only a single zero 
at $(s,q)=(0,0)$, which \cyan{corresponds} neither to a reasonable black hole state nor to 
that of a thermal gas. For $\epsilon=-1$, $\bar\mu$  becomes strictly negative, 
therefore, no HP-like transitions could occur.

\subsection{Clapeyron and Ehrenfest equations for compact TBHs}

When $\epsilon=1$ and $t>t_c$, there is a coexistence phase between the stable small and 
large black holes. The isocharge processes in the coexistence phase are automatically 
isothermal, therefore, according to eq.\eqref{df}, the Helmholtz free energy is 
kept fixed in such processes. 

For simplicity, let us call the stable large and stable small 
black hole phases respectively as phase $A$ and phase $B$. Then, on the coexistence 
isocharge isothermal curves, we have
\begin{align}
-s_{A}\,\rd t+\phi_{A}\,\rd q=-s_{B}\,\rd t+\phi_{B}\,\rd q,
\end{align} 
so the phase curves in the parameter space $(q,t)$ must obey
\begin{align}
\dfrac{\rd q}{\rd t}=\dfrac{s_{A}-s_{B}}{\phi_{A}-\phi_{B}}.
\end{align}
Using the relation $e=f+ts$ and the constancy of $f$ along the phase curve, 
we can rewrite the above equation as
\begin{align}
\dfrac{\rd q}{\rd t}=\dfrac{e_{A}-e_{B}}{t(\phi_{A}-\phi_{B})}. \label{Clap}
\end{align} 
This is the analogy of Clapeyron equation for compact BHs. In order for the above 
equation to be well-defined, the electric potential needs to be discontinuous 
between the two phases. This is a feature required for a first order phase transition. 
In such occasions, there is also a finite jump in $e$, the mean internal energy per 
black hole molecule. If we were looking at the $t-s$ phase transitions from the point of 
view of the whole black hole as \cyan{a} single system, then the jump $e_A-e_B$ would become
a jump in the black hole mass $E_A-E_B=M_A-M_B$, which renders the first order 
phase transitions difficult to understand, because the jump in the mass would break 
the conservation of energy. However, since we are considering finite subsystems, 
the jump $e_A-e_B$ can be attributed to the strong local fluctuations, making it 
sounded physically. More on this point will come out in Section 4.

When $t$ approaches $t_c$, the finite jump for the electric potential becomes smaller 
and smaller, until it vanishes at $t=t_c$. In this case, the right hand side (RHS) of 
eq.\eqref{Clap} becomes ill-defined, and the L’Hospital’s rule needs to be employed 
in order to get a finite value. On this occasion, we have two possible equations,
\begin{align}
\dfrac{\rd q}{\rd t}&=\dfrac{\pfrac{e_{A}}{t}_{q}-\pfrac{e_{B}}{t}_{q}}
{t_c\left[\pfrac{\phi_{A}}{t}_{q}-\pfrac{\phi_{B}}{t}_{q}\right]} ,\label{Ehrenfest1}\\
\dfrac{\rd q}{\rd t}&=\dfrac{\pfrac{s_{A}}{q}_{t}-\pfrac{s_{B}}{q}_{t}}
{\pfrac{\phi_{A}}{q}_{t}-\pfrac{\phi_{B}}{q}_{t}}.\label{Ehrenfest2}
\end{align}
These are the analogues of the Ehrenfest equations. 

The elegant thermodynamic structure of the RPS formalism allows us to inherit the analogous
discussions from ordinary thermodynamics.  Using the Maxwell relation 
\begin{align}
\pfrac{s}{q}_{t}=-\pfrac{\phi}{t}_{q}, 
\end{align}
the denominator in the RHS of eq.\eqref{Ehrenfest1} and numerator in the RHS of  
eq. \eqref{Ehrenfest2} can be related to each other. Moreover, introducing the following 
process parameters
\begin{align}
c_q = t\pfrac{s}{t}_q,\quad
\beta_{q}=\dfrac{1}{\phi}\pfrac{\phi}{t}_{q},\quad
\kappa_{t}=-\dfrac{1}{q}\pfrac{q}{\phi}_{t},
\end{align}
we can express the consistency condition between eqs.\eqref{Ehrenfest1} and 
\eqref{Ehrenfest2} in terms of the following relation among the finite jumps 
of the above process parameters at the critical point,
\begin{align}
\Delta c_{q}|_{t=t_c,q=q_c}
=\Bigg(t\,q\,\phi^{2}\dfrac{\(\Delta\beta_{q}\)^{2}}{\Delta\(\kappa_{t}^{-1}\)}\Bigg)
\Bigg|_{t=t_c,q=q_c,\phi=\phi_c}.
\end{align}
It is evident that $c_q$ is the specific isocharge heat capacity,
and $\beta_{q}$ and $\kappa_{t}$ may be referred to as the isocharge voltage parameter 
and the isothermal discharging parameter, respectively\cite{kong2022restricted}. 
For later use, we present the analytical expression for the 
specific isocharge heat capacity below,
\begin{align}
c_q &=\pfrac{e}{t}_q=\frac{\pfrac{e}{s}_q}{\pfrac{t}{s}_q}
=\frac{(d-3) (d-2) s^{\frac{2(d-3)}{d-2}}\epsilon 
-8 q^2+(d-2) (d-1) s^2 }
{-(d-3)s^{\frac{d-4}{d-2}}\epsilon +\frac{8(2 d-5)}{(d-2)s} q^2+(d-1)s}.
\label{cq}
\end{align}
It can be seen immediately that the case $\epsilon=1$ is distinguished from the 
cases $\epsilon=0, -1$ in that $c_q$ can be divergent in the former case, but not in 
the latter cases provided $d\geq4$. 

\subsection{Local thermodynamic behaviors: a comparison between horizon geometries}

Up till now, our analysis has been carried out in generic spacetime dimensions $d\geq 4$ 
with the exception of the critical point parameters which are calculated in 
concrete dimensions $4\leq d\leq 10$. In this subsection, we wish to make a 
comparison between the detailed thermodynamic behaviors of TBHs with different 
choices of $\epsilon$. For this purpose, we need to create plots of thermodynamic functions 
in various thermodynamic processes. This is a task which is impossible to fulfill 
without specifying a concrete spacetime dimension. Therefore, 
we will be working with $d=4$ throughout this subsection. 
Different choices of $d$ could lead to some quantitative but not
qualitative differences.

The plots for the case $\epsilon=1$ were already presented in 
\cite{gao2021restricted}. However, in order to make comparison to the cases $\epsilon=0,-1$,
we recreated the plots for the case $\epsilon=1$ and put the curves in the same figures 
with the curves with $\epsilon=0,-1$. We believe that this treatment is best for illustrating
the differences in the behaviors of TBHs with different $\epsilon$ values.

\begin{figure}[ht]
\begin{center}
\includegraphics[width=\textwidth]{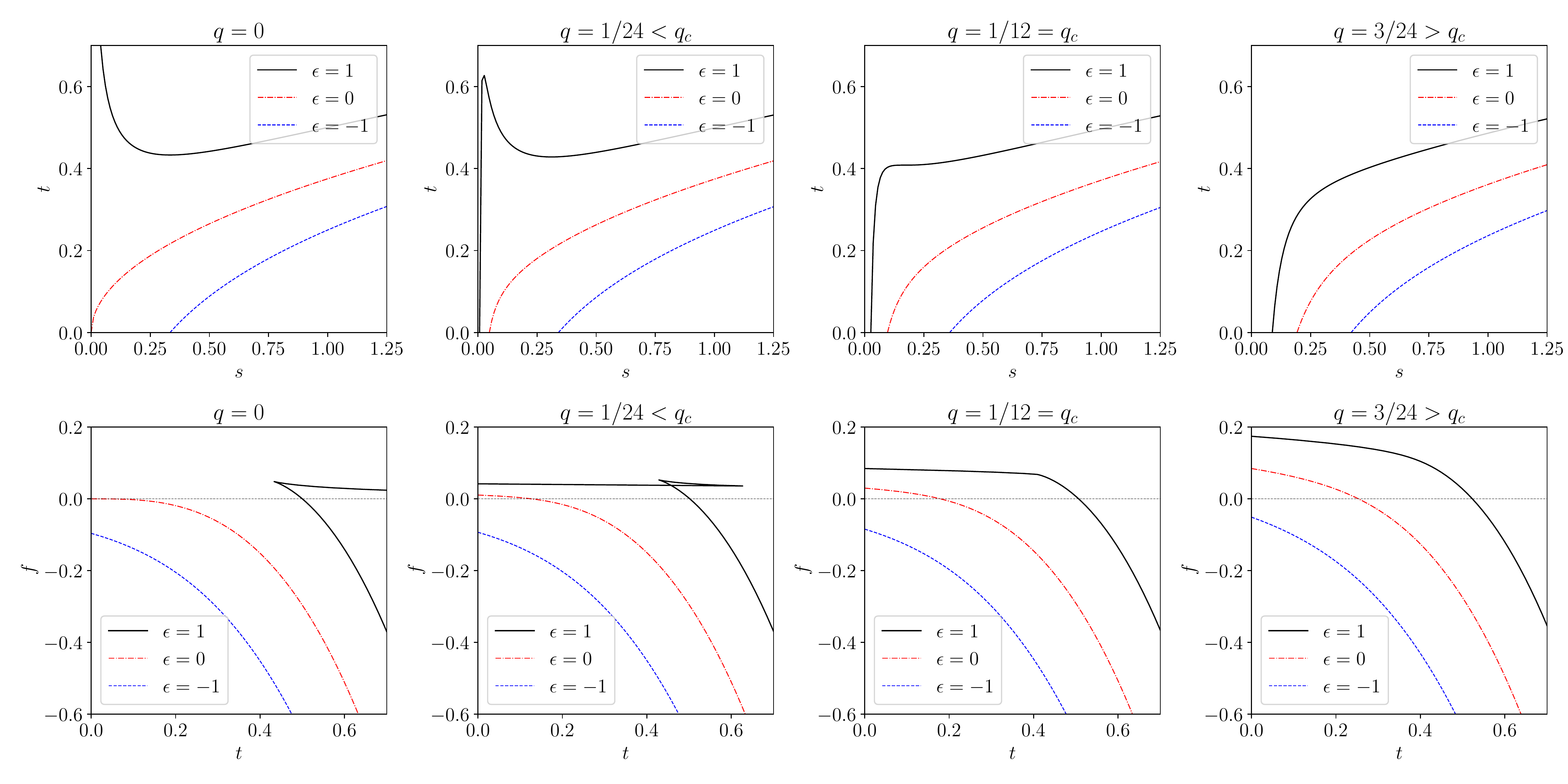}
\caption{The isocharge $t-s$ and $f-t$ curves at different choices of $q$}
\label{fig1}
\end{center}
\end{figure}

\begin{figure}[ht]
\begin{center}
\includegraphics[width=\textwidth]{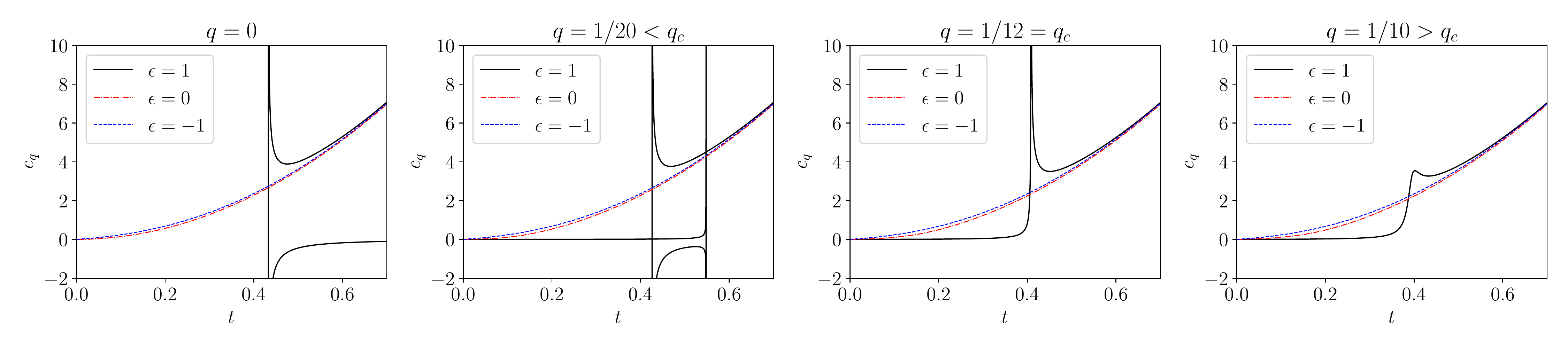} 
\caption{Isocharge specific heat capacity versus temperature}
\label{fig2}
\end{center}
\end{figure}

\begin{figure}[ht]
\begin{center}
\includegraphics[width=\textwidth]{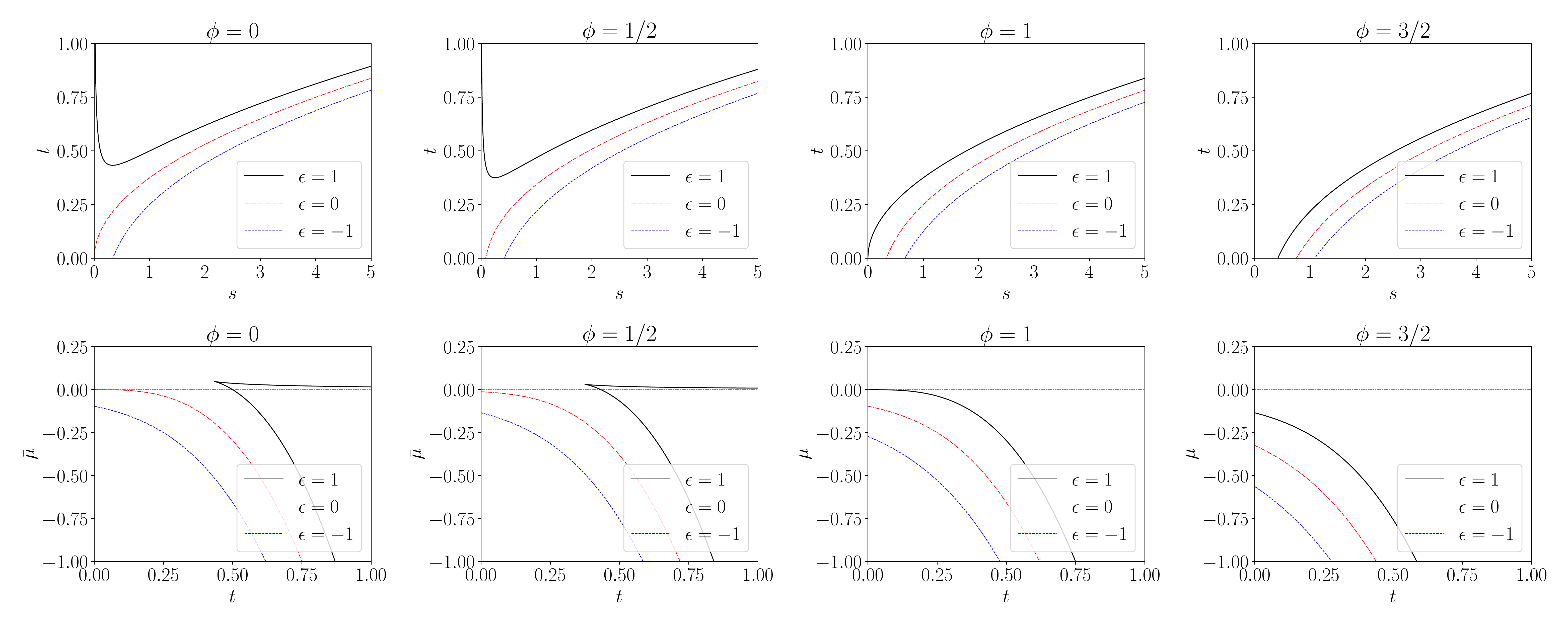}
\caption{The isovoltage $t-s$ and $\bar{\mu}-t$ curves}\label{fig3}
\end{center}
\end{figure}

\begin{figure}[ht]
\begin{center}
\includegraphics[scale=0.5]{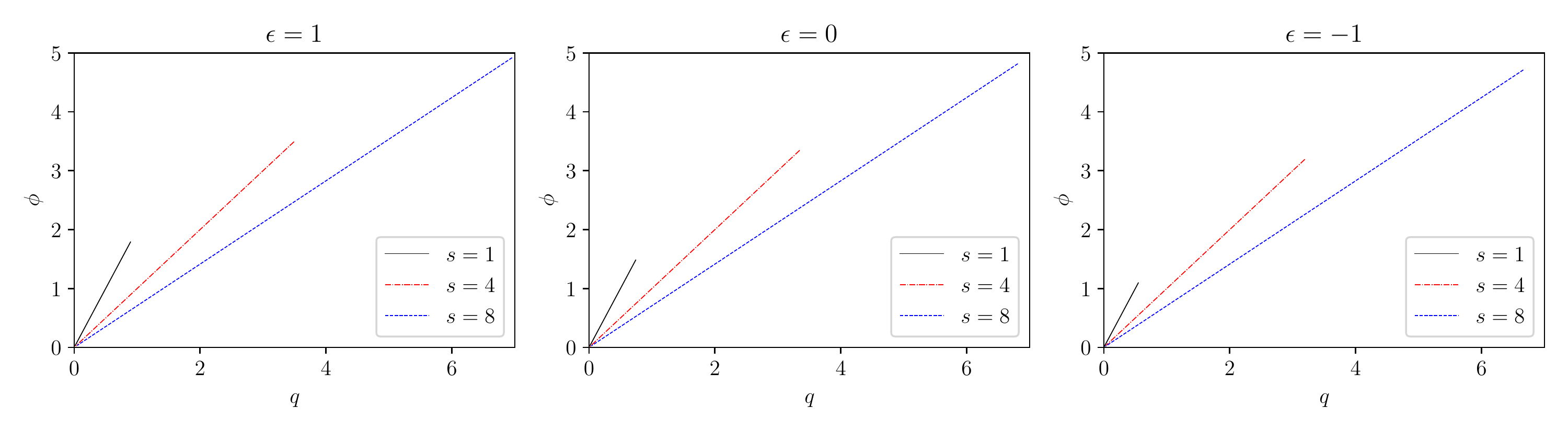}
\caption{The adiabatic $\phi-q$ curves}\label{fig4}
\end{center}
\end{figure}

\begin{figure}[ht]
\begin{center}
\includegraphics[width=\textwidth]{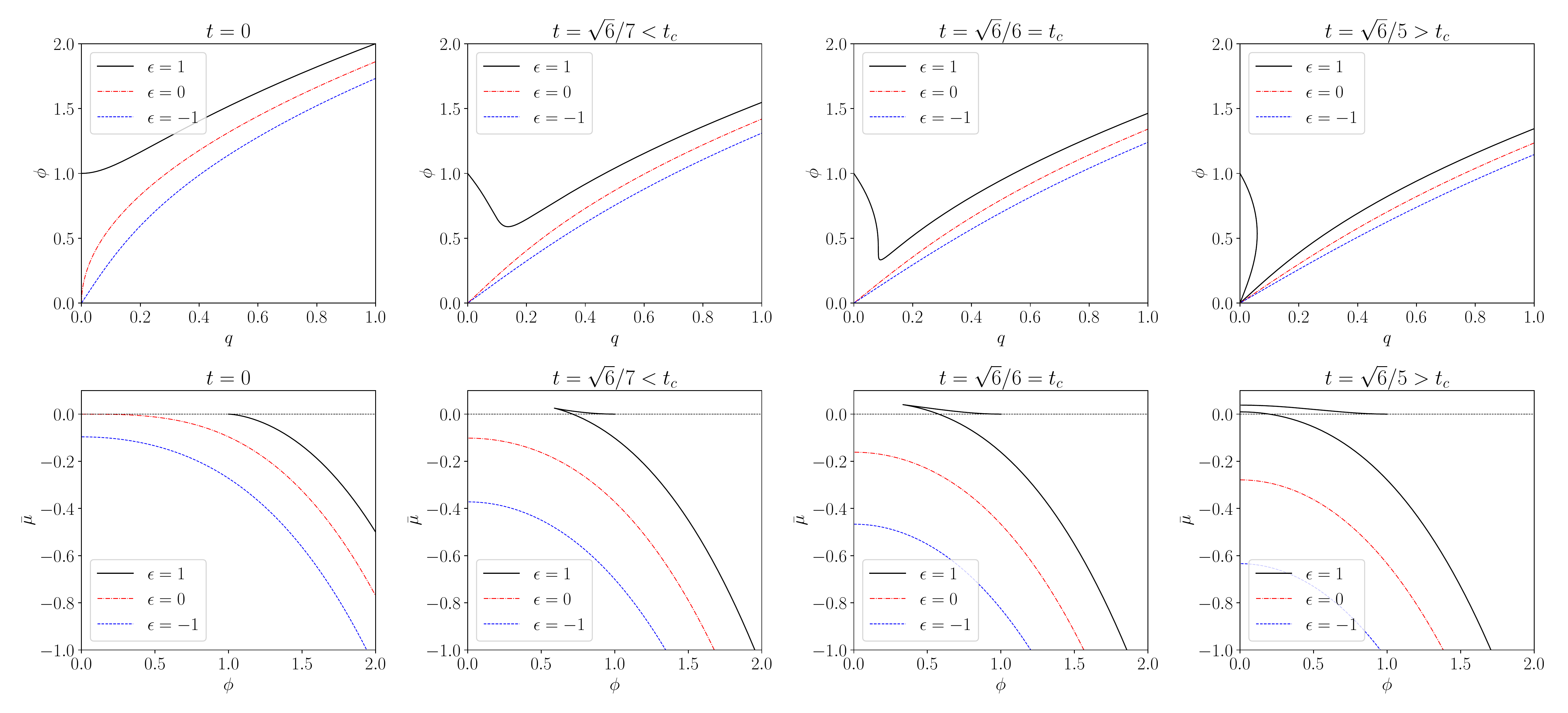}
\caption{The isothermal $\phi-q$ and $\bar{\mu}-\phi$ curves}\label{fig5}
\end{center}
\end{figure}

In $d=4$, eqs.\eqref{e}-\eqref{f} and \eqref{cq} are simplified drastically, yielding
\begin{align}
&e(s,q)=\frac {\epsilon\,s+ 4\,{q}^{2}+s^2 }{4\,\sqrt {s}},
\label{e4d}\\
&t(s,q)=\dfrac{\epsilon\,s-4\,q^{2}+3\,s^{2}}{8\,s^{3/2}},\label{t4d}\\
&\phi(s,q)=\frac{2q}{\sqrt{s}},\label{phi4d}\\
&\bar\mu(s,q)= \frac{\epsilon\,s- 4\,{q}^{2}-s^2}{8\sqrt{s}},\label{barmu4d}\\
&f(s,q)={\frac {\epsilon\,s+ 12\,{q}^{2} - s^2 }{8\sqrt {s}}},\label{f4d}\\
&c_q(s,q)= \frac{2 s \left(\epsilon s -4 q^2+3 s^2\right)}
{-\epsilon s +12 q^2 + 3 s^2}. \label{cq4d}
\end{align}
Notice that the arguments in each of the above equalities are introduced simply 
for reminding the explicit dependence on the independent variables $(s,q)$.
These variables are not necessarily the natural (or adapted) variables for the 
relevant thermodynamic functions. For instance, the adapted independent 
variables for the free energy $f$ should be $(t,q)$, but the 
thermodynamic function of states $f(t,q)$ is not given explicitly but rather 
is provided implicitly by the joint of the relations \eqref{t4d} and \eqref{f4d}.
Of course, one is free to make a change of variable $q\to\phi$ 
in some of the equations listed above by use of 
eq.\eqref{phi4d}. For instance, we can write
\begin{align}
&t(s,\phi)={\frac {\epsilon-{\phi}^{2} +3\,s }{8\sqrt {s}}},
\label{tsphi}\\
&\bar{\mu}(s,\phi)={\frac {\sqrt {s} 
\left( \epsilon-{\phi}^{2}-s\right) }{8}}. \label{musphi}
\end{align}
Once again, the adapted independent variable for $\bar\mu$ regarded as 
a thermodynamic function should be $(t,\phi)$, but \cyan{the} actual relation
$\bar\mu(t,\phi)$ is not presented explicitly but rather is provided implicitly 
by the joint of eqs. \eqref{tsphi} and \eqref{musphi}. 
Moreover, the Bogomol'nyi bound \eqref{bogo1} and \eqref{bogo2} can be solved for $s$, 
which gives a lower bound for each $q$ and $\phi$,
\begin{align}
&s\geq \frac{1}{6}\left(-\epsilon+\sqrt{\epsilon^2+48 q^2}\right),
\\
&s\geq -\frac{\epsilon}{3}+\frac{\phi^2}{3}.
\label{sbound}
\end{align} 
These (in)equalities provide the necessary input for creating the plots given in 
Figs.~\ref{fig1}-\ref{fig5}.

Fig.~\ref{fig1} presents the isocharge $t-s$ and $f-t$ curves of $4d$ TBHs 
in different isocharge processes. The first thing to be noticed in these curves 
is the existence of $t-s$ phase transitions at $0\leq q\leq q_c$ and $t\geq t_c$ 
for the case $\epsilon=1$ (of which the case $q=0$ is an exception in the sense 
that there is a phase transition but no equilibrium temperature). The 
phase transitions at $t>t_c$ are referred to as supercritical. 
Besides the phase transitions which 
occur only for the compact cases, the most remarkable feature to be stressed 
is the existence of states with $t=0, s>0$ in the presence of nonvanishing $q$ 
and all choices of $\epsilon$. Such states are extremal black hole remnants, which indicate 
that charged TBHs cannot be evaporated completely. For $q=0$, 
the state with $t=0$ is unreachable for $\epsilon=1$, but is otherwise 
reachable for $\epsilon=0,-1$. 
For $\epsilon=0$, the state with $q=0, t=0$ corresponds to a
completely evaporated planar black hole ($s=0$), while for $\epsilon=-1$, 
the state with $q=0, t=0$ has a nonvanishing zero point entropy $s$.
Another point which needs to be stressed is that, for $\epsilon=0,-1$, 
$t$ always increases monotonically with $s$ in any isocharge processes, 
signifying that there is only a stable black hole state at any positive 
temperature in these cases.  

The isocharge specific heat capacity $c_q$ corresponding to the above isocharge processes
are presented in Fig.~\ref{fig2}. It can be seen that, for $q=0, 
\,q<q_c,\, q=q_c$ and $q>q_c$, 
the $c_q$ versus $t$ plots have different number of branches for $\epsilon=1$.
In particular, at $q=0$, $c_q$ has two branches, one positive and one negative, 
each corresponds to a stable/unstable compact BH. For $0<q<q_c$, there are three branches 
in the $c_q-t$ plots, two of which are positive and one is negative. 
The two positive branches correspond respectively to the stable small and stable 
large black hole phases, while the negative branch corresponds to the unstable 
medium black hole states. When $q=q_c$, the unstable negative branch disappears, and for 
$q>q_c$, there is only a single positive branch, though the $c_q$ versus $t$ behavior
gradually changes from non-monotonic to monotonic as $q$ increases further. 
In comparison, the $c_q-t$ behaviors for the cases $\epsilon=0,-1$ are much simpler: 
there is always a single, positive-valued branch. 

Besides the different $c_q-t$ behaviors for different choices of $\epsilon$ and different 
values of charge density for the case $\epsilon=1$, there are some other remarkable features
in the $c_q-t$ plots. On the one hand, at high temperatures, the $c_q-t$ curves 
for different choices of $\epsilon$ \cyan{become} more and more close to each other, 
which may signify some universal high temperature \cyan{behaviors}. On the other hand, 
as $t\to 0$, $c_q$ always tends to zero for $\epsilon=0,-1$ and for $\epsilon=1$ with $q=0$,
but at different rates. This indicates that the low temperature behaviors are 
different for different choices of $\epsilon$, and thus are not universal. 
We leave the detailed study on the high and low temperature limits to the next subsection.

Fig.~\ref{fig3} presents the isovoltage $t-s$ and $\bar\mu-t$ curves at zero/nonzero
electric potentials. The $t-s$ and $\bar\mu-t$ curves for the compact BHs 
at $\phi<1$ are distinguished from all other cases, including those for 
compact BHs at $\phi\geq 1$ and for non-compact TBHs at any choices of $\phi$. 
These distinguished curves are either non-monotonic with a minimum for the $t-s$ curves 
or branched with a single zero for the $\bar\mu-t$ curves, which signify the presence of 
HP-like transitions. The curves for the rest cases are all single-valued and monotonic, 
implying the absence of any kind of transitions.

Figs.\ref{fig4} and \ref{fig5} present the adiabatic and isothermal $\phi-q$ curves. 
The adiabatic curves are very easy to understand, because, according to eq.\eqref{phi4d}, 
the electric potential is proportional to the charge in the adiabatic processes. 
There is only one point to be noticed, i.e. every adiabatic $\phi-q$ curve has an 
end point at finite $q$, which is due to the Bogomol'nyi bound. One can see that 
the upper bound for $q$ depends on both $s$ and $\epsilon$. 

The isothermal $\phi-q$ curves are much more subtle, and, 
in order to get better understanding of such processes, we present the curves together 
with the isothermal $\bar\mu-\phi$ plots. It should be mentioned that the analytic 
isothermal $\phi-q$ relations cannot be given explicitly in general, however 
such relationship can be presented in implicit form by considering eqs.\eqref{t4d}
and \eqref{tsphi} jointly, which yields 
\begin{align}
q(s,t)&=\frac{1}{2}\,\sqrt {s}\sqrt {\epsilon+3s-8\,t\sqrt {s}},	
\label{qst}	\\
\phi(s,t)&=\dfrac{1}{a}\,\sqrt {\epsilon+3s-8\,t\sqrt {s}}.
\label{phist}
\end{align}

For compact BHs, the isothermal \cyan{$\phi-q$} relation changes from a single-valued monotonically 
increasing curve to single-valued non-monotonic curve, and then to multivalued 
branched curves, as $t$ increases from $0$ to some supercritical value $t>t_c$. 
The branching occurs only when $t>t_c$. This is actually a manifestation of the 
supercritical $t-s$ phase transitions on isothermal $\phi-q$ curves. Accordingly, 
the isothermal $\bar\mu-\phi$ curve changes from single branched ($t=0$) 
to double branched ($t>0$), with the lower branch crossing the line $\bar\mu=0$, 
signifying HP-like transition. Each branch in the $\bar\mu-\phi$ curve corresponds
to one of the monotonic segment (or branch) on the corresponding $\phi-t$ curve,
and only the segment/branch corresponding to the lower branch of the chemical potential 
is thermodynamically stable. 
For non-compact TBHs, the isothermal $\phi-q$ and $\bar\mu-\phi$ curves
are much simpler and are always single-valued and monotonic. 

Another noticeable feature of the isothermal \cyan{$\phi-q$} curve for compact BHs 
is the existence of a nonzero threshold value for $\phi$ at $t=0, q=0$. 
The analytical value for this threshold 
can be inferred from eqs.\eqref{phist} and \eqref{qst}, which reads
\begin{align}
\phi_{\rm thr}^{2}=\epsilon.
\end{align}
Thus the nonzero threshold for $\phi$ could arise only for $\epsilon=1$. 
In order to understand the origin of the above threshold, it is necessary to 
invert the defining relation for $\phi$ given in eq.\eqref{qN} at $d=4$, which gives
\[
\hat\Phi_{\rm thr}^2=\bfrac{1}{\ell a}^{2} \phi_{\rm thr}^{2} 
=\bfrac{1}{\ell a}^{2}\epsilon.
\]
It is now evident that, for constant $\ell$ and at $a\to \infty$, $\hat\Phi_{\rm thr}^2$
vanishes. We see that the presence of the
threshold is completely a finite size effect, something which is reminiscent to the 
Casimir effect, but now takes effect on the electric potential rather on the energy.

\subsection{High and low temperature limits}

{\bf 1. High temperature limit}

Recently, a correspondence between $d$-dimensional spherically symmetric 
Tangherlini-AdS BHs at high temperatures and quantum phonon gases in $(d-2)$-dimensional
nonmetallic crystals at low temperatures was reported in \cite{kong2022high}. 
The isocharge plots for the $c_q-t$ curves presented in Fig.~\ref{fig2} seem to indicate
that, at high temperatures, the $c_q-t$ behavior does not discriminate 
horizon geometries. Since the plots were created specifically for $d=4$, 
it is necessary to pay a little more efforts to investigate the high temperature behaviors 
analytically in generic spacetime dimensions $d\geq 4$ and for generic choices of 
the parameter $a$. 

At first sight, it may look awkwardly difficult to take the high temperature 
limit for the thermodynamic functions such as $e,f,s,c_q$ {\em etc} given the  
complicated form of eqs.\eqref{e}, \eqref{t}, \eqref{f} and \eqref{cq}. However, the actual 
process in getting the high temperature limit is quite simple. First, by looking at
eq.\eqref{t}, one can see that there are two possible high temperature limits 
for $\epsilon=1$, which correspond to $s\to 0$ and $s\to\infty$ respectively. 
On the other hand, for $\epsilon=0,-1$, there remains only one high temperature 
limit corresponding to $s\to\infty$. Therefore, it is clear that the unified high temperature
limit for all choices of $\epsilon$ takes place at $s\to\infty$, i.e. the high entropy
limit. Then, looking back again at eqs.\eqref{e}, \eqref{t}, \eqref{f} and \eqref{cq}, one 
recognizes that the thermodynamic functions $e,f$ as well as $c_q$ will all 
be dominated by the terms which are independent of $\epsilon$ and $q$. 
This makes it crystal clear that the high temperature limit should not discriminate 
different values of $q$ and $\epsilon$. 
With a few lines of calculation by hand, the high temperature limit behaviors 
of $e,f,s,c_q$ are found to be 
\begin{align}
e&\approx \frac{d-2}{8} \bfrac{t}{t_{\rm BH}}^{d-1} ,\label{eh}\\
f&\approx -\frac{1}{8} \bfrac{t}{t_{\rm BH}}^{d-1} ,\\
s&\approx  \bfrac{t}{t_{\rm BH}}^{d-2} ,\\
c_{q}&\approx (d-2) \bfrac{t}{t_{\rm BH}}^{d-2},\label{cqh}
\end{align}
where 
\[
t_{\rm BH}=\frac{d-1}{8}
\] 
is understood to be corresponding to a characteristic temperature, 
and the high temperature limit means $t\gg t_{\rm BH}$. 
The symbol $\approx$ should be read as 
``approaches as $t\gg t_{\rm BH}$''.

The above power law dependence of thermodynamic quantities
reproduces exactly the results presented in \cite{kong2022high}, and thus coincides with
the low temperature quantum phonon gases in $(d-2)$-dimensional nonmetallic crystals.
The present work generalizes this AdS/phonon gas correspondence to the cases in 
the presence of different charges and horizon geometries. By the way, 
since we can take the liberty of choosing a specific value for the parameter $a$ to 
make the correspondence with phonon gases more precise, i.e. up to numerical coefficients. 
In this way the arbitrariness in choosing the length scale $L$ can be removed.

\noindent {\bf 2. Low temperature limit}

Unlike the high temperature limit which is automatically the high entropy limit,
the temperature for charged TBHs approaches zero at some finite $s$, 
whose value is dependent on $q$. Therefore, the low temperature behaviors for 
charged TBHs can be very complicated in generic dimensions. On the other hand, 
the neutral compact BHs cannot reach the $t=0$ state even asymptotically. Therefore,  
for definiteness, we will consider only the low temperature limits of neutral 
non-compact TBHs in analytic form, and it turns out that the planar and 
hyperbolic TBHs behave differently.

For neutral planar TBHs ($\epsilon=0, q=0$), $t$ becomes a simple power 
in $s$ according to eq.\eqref{t}. Inverting this $t-s$ relation and substituting in 
eqs.\eqref{e}, \eqref{f}, we get
\begin{align}
&s = \bfrac{8t}{d-1}^{d-2} ,\\
&e = \frac{d-2}{8}  \bfrac{8t}{d-1}^{d-1} ,\\
&f = -\frac{1}{8}  \bfrac{8t}{d-1}^{d-1}.
\end{align}
Further, by taking the derivative of $e$ with respect to $t$, we have
\begin{align}
c_q=\pfrac{e}{t}_q = (d-2)\bfrac{8t}{d-1}^{d-2}.
\end{align}
These relations are exact at any temperatures, and the expressions coincide precisely with 
what we have got in the high temperature limit, as shown in eqs.\eqref{eh}-\eqref{cqh}. 
In the low temperature limit ($t\to 0$), the above equations indicate that $s$ and $c_q$  
tend to zero as $t^{d-2}$, while $e$ and $f$ tend to zero as $t^{d-1}$. 

For neutral hyperbolic TBHs ($\epsilon=-1, q=0$), $s$ can be solved explicitly 
from eq.\eqref{t}, yielding
\begin{align}
s=\left(\frac{4t+\sqrt{16t^2+(d-1) (d-3)}}
{d-1}\right)^{d-2}. \label{stn}
\end{align}
Therefore, as $t\to 0$, we have 
\begin{align}
s_0\equiv \lim_{t\to 0} s = \bfrac{d-3}{d-1}^{(d-2)/2}.
\end{align}
Substituting eq.\eqref{stn} into eq.\eqref{e} and taking the first derivative
of the result with respect to $t$, we get, in the limit $t\to 0$, 
the following asymptotic \cyan{behaviors},
\begin{align}
e &\approx |e_0| \(-1+8\,t^2\) ,\\
c_{q} &\approx 16\,|e_0|\, t,
\end{align}
where
\begin{align}
e_0\equiv \lim_{t\rightarrow 0}e =-\dfrac{d-2}{4(d-1)}\bfrac{d-3}{d-1}^{(d-3)/2}.
\end{align}
The Helmholtz free energy has precisely the same zero point value as the internal energy,
\[
f_0\equiv \lim_{t\rightarrow 0}f=e_0,
\]
however, above absolute zero, $f$ develops a lowest order correction term 
which is linearly dependent on $t$. 

For charged TBHs with generic horizon geometries, the analytic analysis for the 
low temperature limits is impossible to be carried out, therefore we resorted to 
numeric methods. To our surprise, for $4\leq d\leq 10$ and generic $q>0$, 
all charged TBHs with generic horizon geometries have similar low temperature 
behaviors, e.g. the presence of non-vanishing zero point values $e_0=f_0\neq0$ which depend
on $q$ and the linear behavior $c_q\propto t$ for the specific heat capacity. 

The above results indicate that, at low temperatures, all charged TBHs behave like 
strongly degenerated electron gases. 
The neutral planar case is an exception to this generic behavior, probably 
because the neutral condition in this case happens to have destroyed the coefficients of 
all terms in the series expansions for $e$ and $f$ besides the $t^{d-1}$ terms.

It is remarkable that, besides the presence of a nonvanishing zero point entropy $s_0$ 
and the negative zero point energy $e_0$, the low temperature asymptotic behaviors 
of the charged TBHs and the neutral hyperbolic TBHs are very similar to that of 
the strongly degenerated 
electron gas, especially regarding the $e\sim e_0+\alpha_1 t^2$ behavior for the 
internal energy and the linear behavior $c_q\sim\alpha_2 t$ for the 
specific heat capacity. As $t$ increases from $0$ to $t\gg t_{\rm BH}$, 
the behaviors of the charged TBHs and the neutral hyperbolic TBHs smoothly 
interpolate between that of strongly degenerated electron gas and quantum phonon gas. 
This result is another surprising discovery of this work in addition to the proof of 
universal high temperature behaviors for TBHs with different horizon geometries.

\section{TBHs in $4d$ conformal gravity}

\subsection{Thermodynamic Structure}

The second model which we would like to consider in this work is the 4-dimensional 
conformal gravity. The thermodynamics of the charged spherically symmetric AdS BHs 
has been discussed in \cite{kong2022restricted} under the RPS formalism. 
The reason why we look back at this model is because the thermodynamic behavior of the 
non-compact TBHs in this model looks quite different, or even physically 
ill-posed. This opens a new possibility for testing the physical viability 
of various gravity models from the thermodynamic perspective.

Before dwelling into the details of the thermodynamic behavior, we need to 
recall some basic data for this specific model of gravity.
The action with the electromagnetic field is given as
\begin{align}
S=\alpha\int \rd^4x \sqrt{-g}\left(\frac{1}{2}C^{\mu\nu\rho\sigma}
C_{\mu\nu\rho\sigma} +\frac{1}{3} F^{\mu\nu}F_{\mu\nu}\right),
\label{conaction}
\end{align}
where we assume $\alpha=\frac{L^{2}}{16\pi G}$ in order to keep $\alpha$ dimensionless 
as it should \cite{kong2022restricted}. The static charged AdS black hole 
solution for this model can be found in \cite{li2013fermi}, with the metric
\begin{align}
&\mathrm{d} s^2=-f(r)\mathrm{d}t^2+\frac{\mathrm{d}r^2}{f(r)}+r^2\mathrm{d}
\Omega_{2,\epsilon}^2, \label{conmetric}\\
&f(r)=-\frac{1}{3}\Lambda r^2+c_1 r+c_0+\frac{d}{r},\label{confr}
\end{align}
and the Maxwell field
\begin{align}
A=-\frac{Q}{r}\mathrm{d}t.
\end{align}
The parameter $\epsilon=1, 0, -1$ is the same as in the previous case, and 
the other parameters $Q, c_0, c_1, d, \Lambda$ 
are all integration constants which need to obey an additional constraint
\begin{align}
  3c_1d+\epsilon^2+Q^2=c_0^2,
\label{conrelation}
\end{align}
and, for the purpose to have a reasonable vacuum configuration ($c_1=d=Q=0$) 
and also for the consistency of the RPS description for this model, we need to take
$c_0=\epsilon$.

The charged AdS TBH is located at one of the real positive root $r_0$ of the equation
\begin{align}
f(r)=0. \label{evh}
\end{align}
The basic thermodynamic quantities for these TBHs can be \cyan{found} in \cite{li2013fermi}, 
\begin{align}
&E=\dfrac{\omega_{2,\epsilon}\alpha\(c_{0}-\epsilon\)\(\Lambda r_{0}^{2}-3 c_{0}\)}
{72\pi r_{0}}+\dfrac{\omega_{2,\epsilon} \alpha d\(2 \Lambda r_{0}^{2}-c_{0}+\epsilon\)}
{24\pi r_{0}^{2}} ,\\
&S=\dfrac{\omega_{2,\epsilon} \alpha\(\epsilon r_{0}-c_{0} r_{0}- 3 d\)}{6 r_{0}} ,\quad\quad 
T=-\dfrac{\Lambda r_{0}^{3}+3 c_{0} r_{0} + 6 d}{12 \pi r_{0}^{2}} ,\\
&{\hat Q}=\dfrac{\omega_{2,\epsilon} \alpha Q}{12 \pi},\quad\quad\quad\quad 
\hat\Phi=-\dfrac{Q}{r_{0}} ,
\end{align}
where $\omega_{2,\epsilon}$ represents the area of the 2-dimensional submanifold designated 
by the line element $\rd\Omega_{2,\epsilon}$. The opposite sign 
between ${\hat Q}$ and $\Phi$ is specific to this model, which may be originated from the 
unusual sign convention in the action for the Maxwell field.

In the RPS formalism, we introduce 
\begin{align}
N=\dfrac{L^{2}\omega_{2,\epsilon}}{G}=16\pi\alpha\omega_{2,\epsilon},\quad\quad 
\mu=\dfrac{GTI_{E}}{L^{2} \omega_{2,\epsilon}}
=\dfrac{2\left[d\(3\epsilon - r^{2}\Lambda\)
+2 \epsilon r\(c_{0}-\epsilon \)\right]}{3\,r^{2}}.
\end{align}
These quantities guarantee the correctness of the Euler relation, the first law and the 
Gibbs-Duhem relation as given in eqs.\eqref{Euler}-\eqref{GDR}. 
Please notice that, for non-compact TBHs, 
$\omega_{2,\epsilon}$ diverges. Therefore, to actually make sense of the thermodynamics for 
the non-compact cases, the introduction of finite subsystems is unavoidable.

\subsection{Local thermodynamic relations and high temperature limit}

We proceed in analogy to the case of Einstein-Maxwell theory. The necessary 
local thermodynamic variables are defined as
\begin{align}
e&=\dfrac{2\pi\ell E}{N},\qquad s\equiv \frac{S}{N},\qquad q\equiv \frac{{\hat Q}}{N},\\
&\bar\mu \equiv 2\pi\ell\mu,\quad t\equiv 2\pi\ell T,\qquad \phi\equiv 2\pi\ell \hat\Phi,
\end{align}
and also
\begin{align}
f=e-ts.
\end{align}
Taking $(s,q)$ as independent variables, the other local thermodynamic functions 
are found to be
\begin{align}
&e=\sqrt{s\left(-s \epsilon-384 \pi^3 q^2+32 \pi s^2\right)}, \\
&t=\dfrac{1}{\sqrt{s}}\frac{-s \epsilon-192 \pi^3 q^2+48 \pi s^2}
{\sqrt{-s \epsilon-384 \pi^3 q^2+32 \pi s^{2}}},\\
&\phi=-\frac{384 \pi^3 q\,\sqrt{s}}{\sqrt{-\epsilon s-384 \pi^3 q^2+32 \pi s^{2}}},\\
&\bar{\mu}= -\frac{16 \pi\,\sqrt{s}\left(s^2-12 \pi^2 q^2\right)}
{\sqrt{-s \epsilon-384 \pi^3 q^2+32 \pi s^{2}}}.
\end{align}
Using the first two of these \cyan{formulae}, the specific isocharge heat capacity 
is calculated to be
\begin{align}
c_q=-\frac{s \left(192 \pi ^3 q^2-48 \pi  s^2+s \epsilon \right) 
\left(384 \pi ^3 q^2-32 \pi s^2+s \epsilon \right)}
{32 \pi  \left(1152 \pi ^5 q^4+576 \pi ^3 q^2 s^2-24 \pi  s^4
+s^3 \epsilon \right)}.
\end{align}
Just like in the case of Einstein-Maxwell theory, it suffices to consider only 
the cases with $q\geq 0$ (and thus $\phi\leq 0$).

In order to ensure the above functions to be real-valued, the expression under the 
square root needs to be non-negative. This gives a bound
\begin{align}
s\geq \dfrac{\epsilon+\sqrt{12\,(64\,\pi^{2}\,q)^{2}+\epsilon^{2}}}{64\pi}
\label{sboundcfg}
\end{align}
for each $q$. It happens that the above bound automatically ensures the non-negativity
of $t$, thus there is no need to consider the Bogomol'nyi bound in this case.

The above explicit form for the thermodynamic functions allows us to consider 
the common high temperature limit $t\to\infty$ for all horizon geometries, 
which is also the large entropy limit.
The results read
\begin{align}
\lim_{t\rightarrow+\infty} e&=\dfrac{t^{3}}{108\pi} ,\quad\quad
\lim_{t\rightarrow+\infty} f=-\dfrac{t^{3}}{216\pi} ,\\
\lim_{t\rightarrow+\infty} s&=\dfrac{t^{2}}{72\pi} ,\quad\quad\,\,\,
\lim_{t\rightarrow+\infty} c_{q}=\dfrac{t^{2}}{36\pi}.
\end{align}
It is worth to notice that these results are independent of $q$ and $\epsilon$, and 
the same power law dependence on $t$ as in eqs.\eqref{eh}-\eqref{cqh} has appeared
once again, further supporting our conjecture about the universality of the 
AdS/phonon gas correspondence. 

Please be reminded that, for neutral planar TBHs in conformal gravity, the above 
power law dependence on $t$ of the thermodynamic quantities are exact, irrespective of
the high temperature limit. Thus the neutral planar TBHs behave like $2d$ phonon gases
even in the low temperature limit. Besides the neutral planar case, the TBHs in 
conformal gravity do not admit a low temperature limit, which is due to the fact 
that the bound \eqref{sboundcfg} is actually tighter than the Bogomil'nyi bound, 
making the states with $t=0$ but $s\neq 0, q\neq0$ inaccessible.
This leaves no more room for studying the low temperature limit.

\subsection{Description of thermodynamic processes}

A significant difference of conformal gravity from Einstein-Maxwell theory is the 
absence of isocharge $t-s$ criticality. This makes the isocharge processes for
TBHs in conformal gravity much simpler. 

\begin{figure}[ht]
\begin{center}
\includegraphics[width=.7\textwidth]{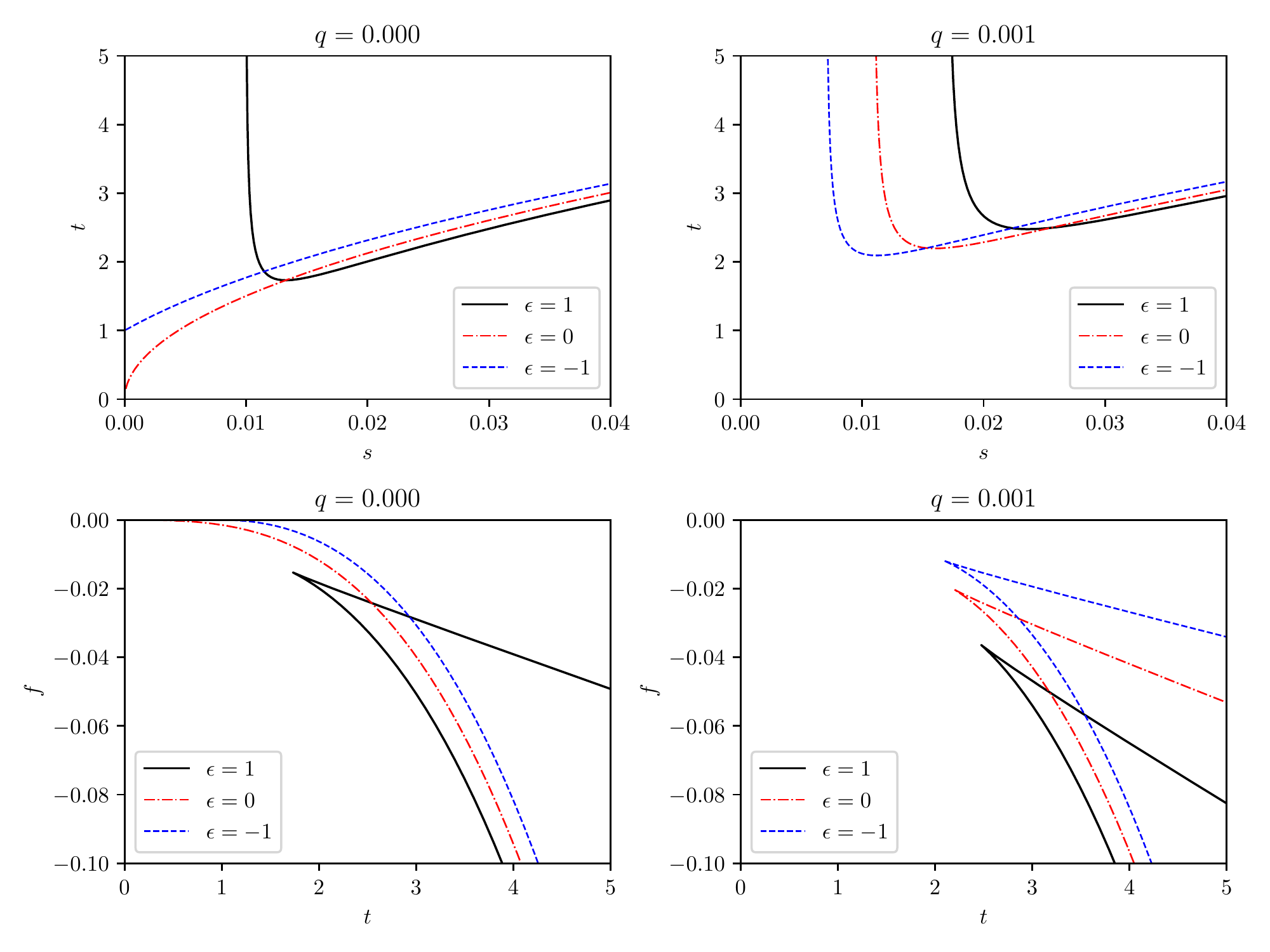}
\caption{The isocharge $t-s$ and $f-t$ curves for TBHs in conformal gravity}\label{fig6}
\end{center}
\end{figure}

\begin{figure}[ht]
\begin{center}
\includegraphics[width=.8\textwidth]{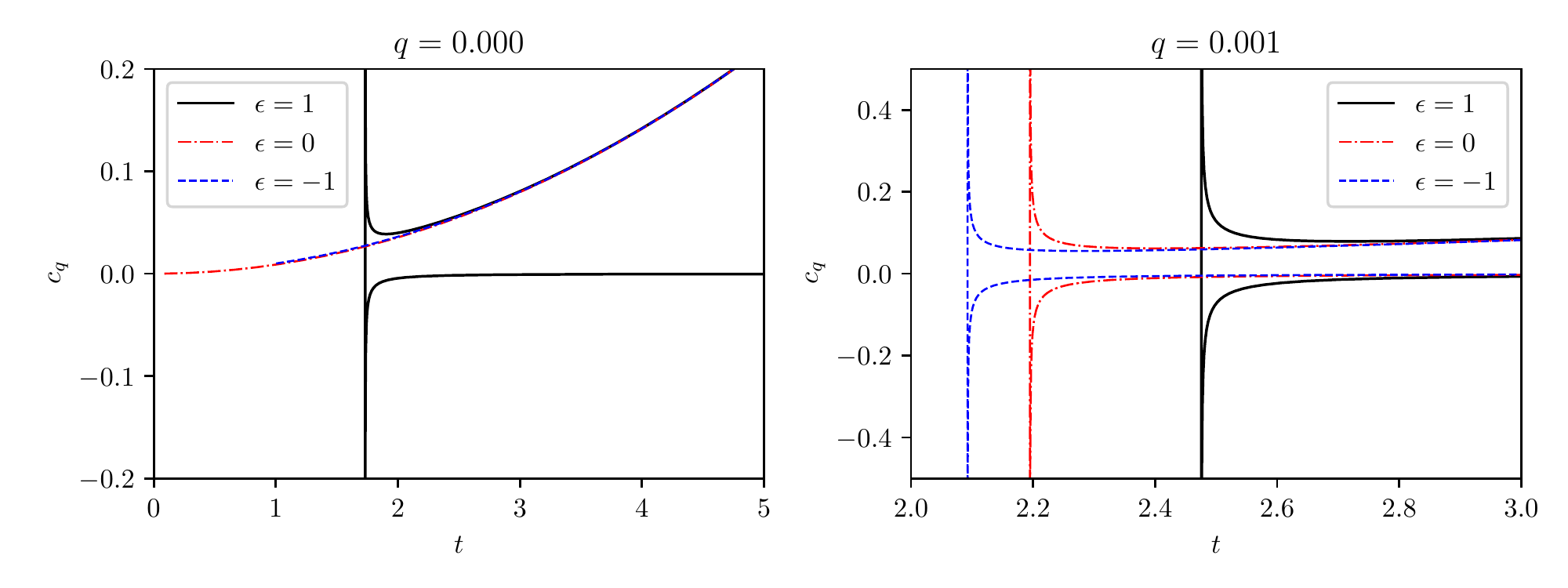} 
\caption{The isocharge specific heat capacity versus temperature for conformal gravity}\label{fig7}
\end{center}
\end{figure}

\begin{figure}[h!]
\begin{center}
\includegraphics[width=.8\textwidth]{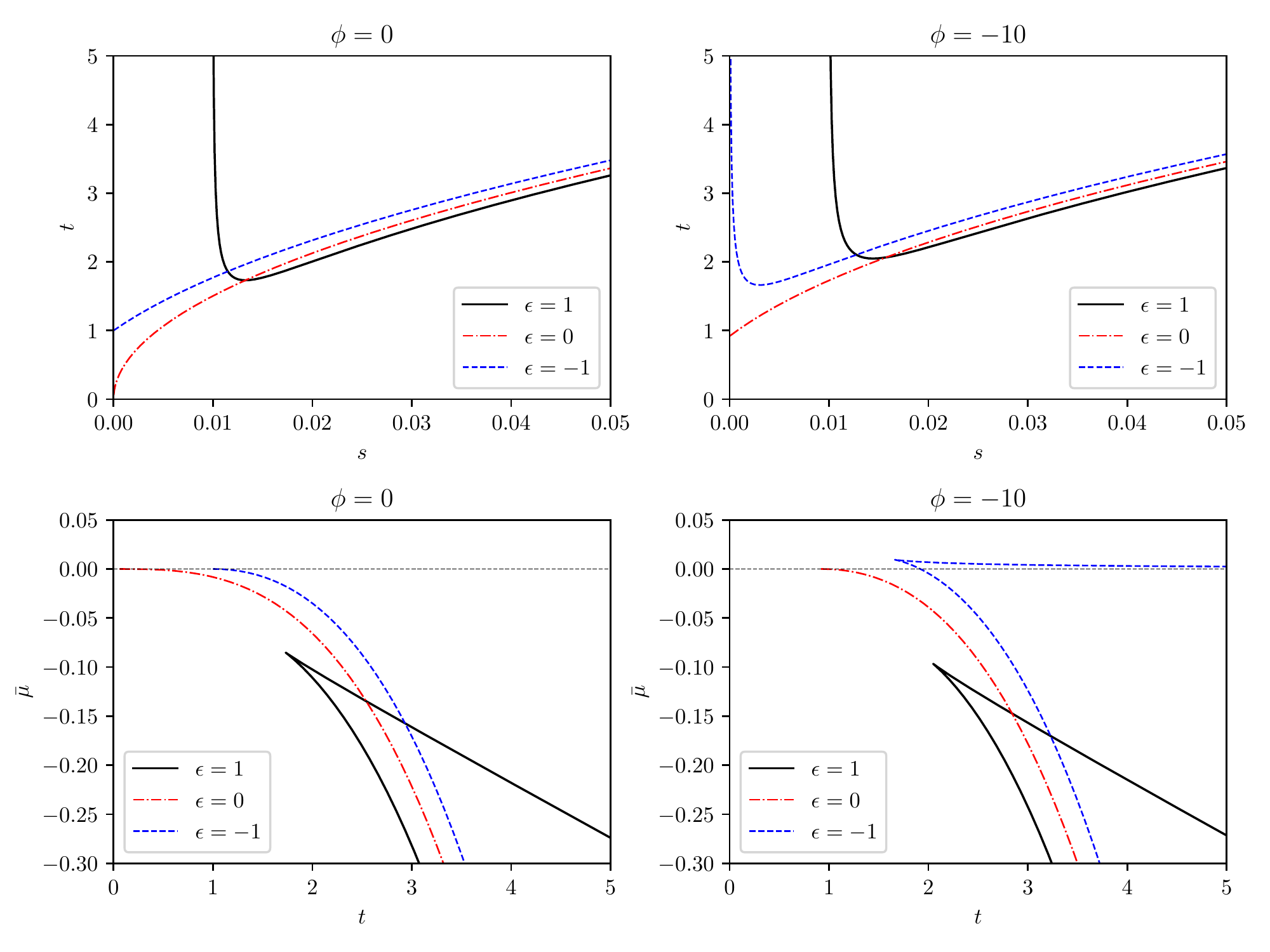}
\caption{The isovoltage $t-s$ and $\bar{\mu}-t$ curves for TBHs in conformal gravity}
\label{fig8}
\end{center}
\end{figure}

The isocharge $t-s$ and $f-t$ plots for neutral and charged TBHs 
in $4d$ conformal gravity are presented in Fig.~\ref{fig6}. The solid curves 
correspond to compact spherically symmetric BHs, which were first obtained in 
\cite{kong2022restricted} and are reproduced here in order to make comparison 
to the non-compact TBHs. Besides the branched/un-branched behaviors for the 
Helmholtz free energy, there is a stunning difference between the neutral hyperbolic 
BH and all the other cases. The point lies in that the lowest entropy state 
for the neutral hyperbolic BH is located at $(s,t)=(0,1)$. This is quite unusual, 
because it is commonly believed that a temperature exists only for thermal systems with 
nonzero entropy. A state with zero entropy but nonzero temperature is beyond 
the usual thermodynamic understandings, and this may be a signature for the illposed-ness 
of conformal gravity, or of its hyperbolic black hole solution, 
although it has been argued that this model is on-shell equivalent to
Einstein gravity with a negative cosmological constant \cite{Maldacena:2011mk}. 
In this regard, please be reminded that, in Einstein gravity with a 
negative cosmological constant, the hyperbolic black holes do not have 
similar problems. This may be the first example 
case for testing the viability of gravity models or some of their 
solutions from thermodynamic perspective.

Now let us ignore the above problem for the neutral hyperbolic BH and proceed the analysis
by presenting other plots. Fig.~\ref{fig7} presents the plots of isocharge specific 
heat capacity versus temperature. These plots look similar to those for the neutral TBHs
in Einstein-Maxwell gravity as presented in the left-most figure in Fig.~\ref{fig2}. 
However, there is a noticeable difference, i.e. the curve corresponding to  
the neutral hyperbolic BH does not extend to the origin on the $t-c_q$ plane. 

The isovoltage $t-s$ and $\bar{\mu}-t$ curves are presented in Fig.~\ref{fig8}, 
which \cyan{look} similar to the isocharge plots given in Fig.~\ref{fig6}. 
As already pointed out in \cite{kong2022restricted}, for spherical BHs in conformal 
gravity, there is no HP-like transitions because the chemical potential is strictly 
negative. One surprising fact which comes with Fig.~\ref{fig8} is that,
for charged hyperbolic BHs in conformal gravity, the HP-like transition could occur. 
This behavior is in sharp contrast to the cases in Einstein-Maxwell theory. 
As can be inferred \cyan{from} Fig.~\ref{fig3}, the HP-like transition could occur only 
for compact BHs in Einstein-Maxwell theory. Now in conformal gravity, the situation 
seems to be inverted. Another important point to observe in Fig.~\ref{fig8} 
is the existence of states $(s=0,t>0)$. Such states already show up in Fig.~\ref{fig6} 
for neutral hyperbolic BHs, but now reappear for both planar and hyperbolic 
TBHs. The appearance of such unphysical states on the isocharge and isovoltage $t-s$ 
curves drives us to raise concerns about the consistency of conformal gravity 
as a candidate for a physically viable alternative theory of gravity, 
or, at least, question the validity of non-compact TBHs as physically viable solutions.
It may be reasonable to impose constraints over the integration constants
by thermodynamic considerations, e.g. to set $\epsilon=1$. 

\begin{figure}[h]
\begin{center}
\includegraphics[width=.8\textwidth]{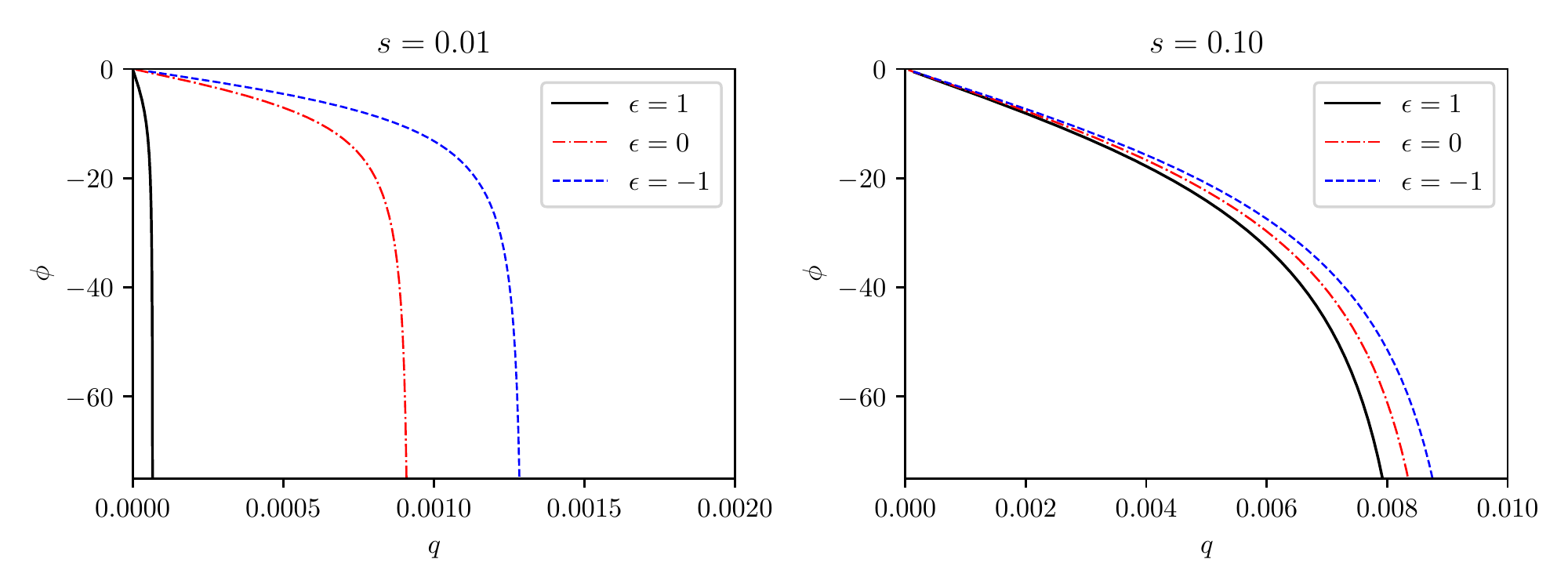}
\caption{The adiabatic $\phi-q$ curves for TBHs in conformal gravity}\label{fig9}
\end{center}
\end{figure}
\begin{figure}[ht]
\begin{center}
\includegraphics[width=.99\textwidth]{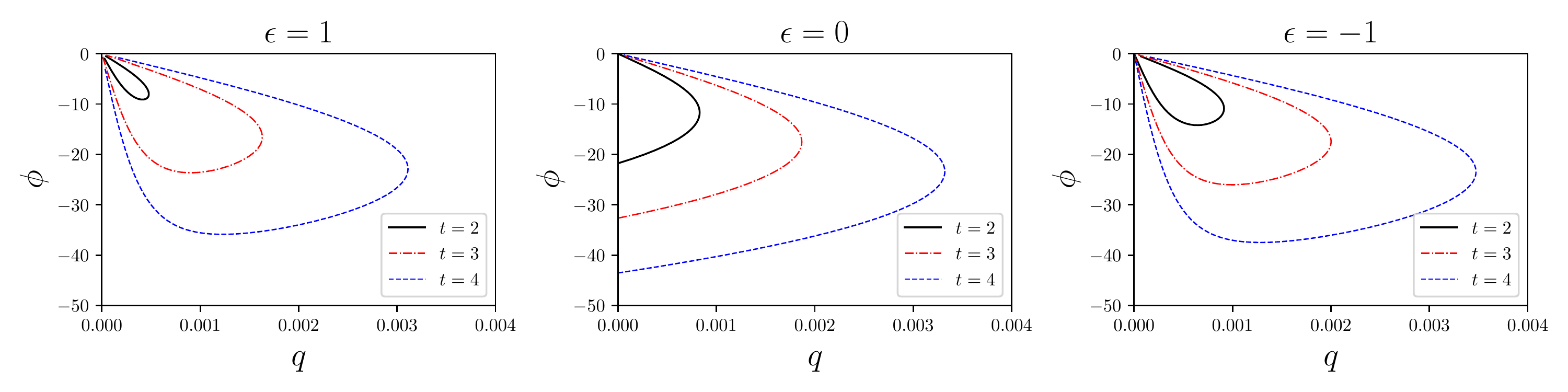}
\caption{\cyan{The isothermal $\phi-q$ }curves for TBHs in conformal gravity}
\label{fig10}
\end{center}
\end{figure}

\begin{figure}[h!]
\begin{center}
\includegraphics[width=.7\textwidth]{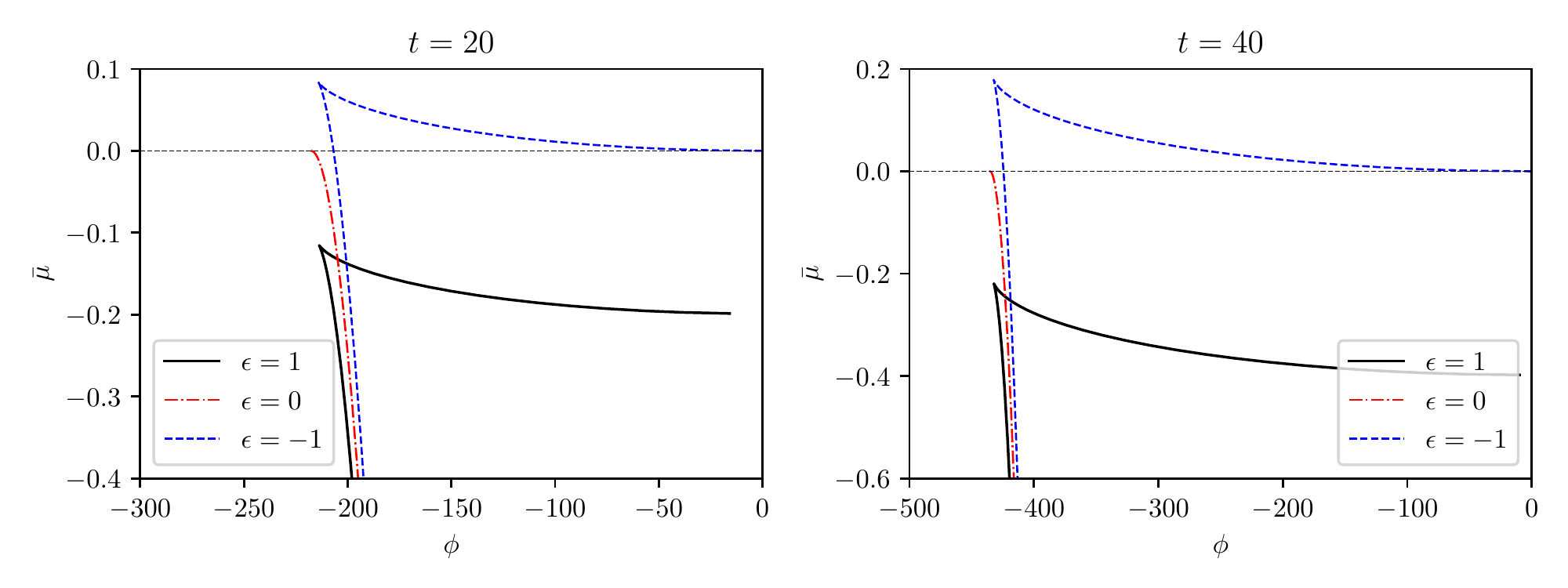}
\caption{The isothermal $\bar{\mu}-\phi$ curves for TBHs in conformal gravity}
\label{fig11}
\end{center}
\end{figure}

For completeness, we also present the the 
$\phi-q$ curves in adiabatic and isothermal processes in Fig.~\ref{fig9} 
and Fig.~\ref{fig10}. There is no significant differences between the 
adiabatic $\phi-q$ curves for TBHs with different horizon geometries, however 
the isothermal \cyan{$\phi-q$} curves for planar TBHs are distinguished from the 
spherical and hyperbolic cases.

Finally, we present the isothermal $\bar{\mu}-\phi$ curves for TBHs in conformal
gravity in Fig.~\ref{fig11}, which may be used in conjunction with the 
isovoltage $\bar{\mu}-t$ curves as presented in Fig.~\ref{fig8} for illustrating 
the presence/absence of HP-like transitions in conformal gravity. 

\section{Fluctuations in subsystems}

The thermodynamic fluctuations for black holes have been considered by different authors 
in the past, either by use of Smoluchowski formula or in terms of thermodynamic geometry
\cite{aaman2003geometry,aaman2006geometry,sahay2010thermodynamic,zhang2015phase,
wei2017critical,wei2019ruppeiner,xu2020diagnosis,xu2020ruppeiner,Wc2022,
peca1999thermodynamics}. 
All previous works on this subject have treated the black hole as a whole system, 
without subdividing it into subsystems. Please be reminded that the Smoluchowski
formula is derived for a finite thermodynamic system by immersing it 
in a large heat reservoir, which is in turn considered to be an isolated system. 
For a black hole as a whole system, the role of reservoir remains obscure. 
On the other hand, in approaches with the use of thermodynamic geometry 
such as the Ruppeiner geometry\cite{ruppeiner1981application,ruppeiner1995riemannian}, 
one looks for the divergent points of the Ricci scalar associated with the Ruppeiner 
metric (which is defined by the matrix of second derivatives of the entropy as a 
function of states) for potential phase transition behaviors. Why Riemannian geometry 
plays any role in the space of macro states for thermodynamic systems 
seems to have never been explained sufficiently clear. In particular, the 
meaning of general coordinate transformations on the space of macro states 
\cyan{remains} poorly understood, especially when Euler homogeneity has to be taken 
into account.

In this section, we shall discuss the thermodynamic fluctuations for TBHs using some 
of the ideas developed in Section 2. Our standing point is still based on subsystems,
and the fluctuations which we discuss will be referred to as local fluctuations. 
Some of the conceptual issues in the previous treatment of black hole thermodynamic
fluctuations will be clarified on the fly with our discussions.

\subsection{The approach with Smoluchowski formula}

The concept of subsystems introduced in Section 2 allows us to consider the local 
thermal fluctuations in the subsystems using standard thermodynamic methods.

The major conceptual issue in treating the black hole fluctuations using 
Smoluchowski formula lies as follows. On the one hand, the derivation of 
Smoluchowski formula requires a heat reservoir containing a large number 
of particles (i.e. $N\to\infty$), so that the relative fluctuations becomes negligible. 
On the other hand, for black holes regarded as a whole system, it is hard to find 
an appropriate reservoir to carry out the process for deriving Smoluchowski formula. 
One might imagine to immerse the black hole in an external thermal environment 
whose scale is much larger than the black hole itself, and take the environment 
as reservoir. However, this picture is clearly incorrect, because the black hole 
cannot stay in equilibrium with the environment. Rather, it may
absorb matter and energy from the environment and evolves into a larger black hole 
at a different thermodynamic state. 

The correct way to deal with the problem is to take the black hole itself as the 
reservoir, and consider the local fluctuations for its finite subsystems. 
For non-compact TBHs, the horizon area is already infinitely large, 
so are the additive quantities such as the total internal energy, entropy, 
electric charge and the number of black hole molecules contained therein. 
Therefore there is no problem in considering such black holes as reservoirs. 
While for compact BHs, the horizon area is finite, and so are the additive quantities.
This may raise concerns about the validity for taking 
such black holes as heat reservoirs. The way out is to consider the large 
$L$ or small $G$
limit, in which case the number $N$ of black hole molecules blows up, 
and it becomes more reasonable to take such black holes as reservoirs. 

According to the above reasoning, we can now always take the whole black hole 
as a heat reservoir. All we need is to keep in mind that the number 
$N$ of black hole molecules must be considered to be very large. Taking a subsystem 
which corresponds to a fixed small area on the event horizon (small in the sense 
$\s\ll S$, but still large enough to make $\n\gg 1$), the corresponding 
additive quantities of the subsystem must obey
\[
\e\ll E,\quad 1\ll\n\ll N,\quad \s\ll S,\quad \q \ll Q. 
\]
The corresponding thermodynamic identity for the subsystem can be written as
\begin{align*}
\rd\e=T\rd\s+\hat\Phi\rd\q +\mu \rd\n. 
\end{align*}
In accordance with the dimensionless 
notations introduced in Sections 2 and 3, we also introduce the dimensionless additive 
quantities for the subsystem,
\begin{align}
\en = \n e,\quad \sn=\n s, \quad \qn = \n q. \label{esqn}
\end{align}
Then, the thermodynamic identity for the subsystem can be rewritten as
\begin{align*}
\rd\en=t\rd\sn+\phi\rd\qn +\bar\mu \rd\n.
\end{align*}
When $\rd \n =0$, the subsystem is closed. 

Following the standard procedure, we can get the Smoluchowski
formula for the probability density corresponding to the fluctuation configuration 
$( \delta t,\delta\sn,\delta\phi,\delta\qn )$ in a small closed subsystem,
\begin{align}
\p = A\exp\(-\frac{\delta t\delta\sn+\delta\phi\delta\qn}{2t}\), \label{prd}
\end{align}
where $A$ is an appropriate normalization constant. 

In order to make use of the above 
distribution in the evaluation for the relative fluctuations, one needs to be aware 
of the fact that, among the four fluctuation quantities, only two are independent. 
As an example case, we can take $(\delta \sn,\delta\qn)$ as independent fluctuations, 
so that the other two fluctuations can be expressed as
\begin{align*}
\delta t&=\pfrac{t}{\sn}_{\qn}\delta \sn+\pfrac{t}{\qn}_{\sn}\delta\qn,\quad
\delta\phi =\pfrac{\phi}{\sn}_{\qn}\delta \sn+\pfrac{\phi}{\qn}_{\sn}\delta\qn.
\end{align*}
With the aid of the Maxwell relation
\begin{align*}
\pfrac{t}{\qn}_{\sn}&=\pfrac{\phi}{\sn}_{\qn} ,
\end{align*}
the probability density \eqref{prd} becomes
\begin{align}
\p=A\exp\left\{-\dfrac{1}{2\,t}\left[\pfrac{t}{\sn}_{\qn}\(\delta \sn\)^{2}
+\pfrac{\phi}{\qn}_{\sn}\(\delta \qn\)^{2}
+2\pfrac{t}{\qn}_{\sn}\delta\qn\delta\sn \right]\right\}. 
\label{p1}
\end{align}
Using this probability density, it is straightforward to calculate the 
following mean-square (relative) fluctuation,
\begin{align}
&\la (\delta \sn)^2 \ra = t\pfrac{\sn}{t}_{\qn}=\n c_q,\quad
\la\(\dfrac{\delta \sn}{\sn}\)^{2}\ra 
=\frac{\la (\delta \sn)^2 \ra}{\sn^2} 
=\frac{c_q}{\n s^2} \propto \dfrac{1}{\n}.
\end{align}
Likewise, we have
\begin{align}
\la\(\delta \qn\)^{2}\ra 
=\dfrac{t\pfrac{t}{\sn}_{\qn}}
{\pfrac{t}{\sn}_{\qn}\pfrac{\phi}{\qn}_{\sn}-\left[\pfrac{t}{\qn}_{\sn}\right]^2}, \quad
\la\(\dfrac{\delta \qn}{\qn}\)^{2}\ra 
=\dfrac{\la\(\delta \qn\)^{2}\ra }{\qn^{2}} \propto \dfrac{1}{\n},
\label{fluctq}
\end{align}
and there is a nontrivial correlation between the entropy and charge fluctuations,
\begin{align}
\la\delta\sn\delta\qn\ra=-\dfrac{t\pfrac{t}{\qn}_{\sn}}
{\pfrac{t}{\sn}_{\qn}\pfrac{\phi}{\qn}_{s}-\left[\pfrac{t}{\qn}_{\sn}\right]^2}.
\label{correlqs}
\end{align}

Alternatively, taking $(\delta\sn,\delta\phi)$ as independent fluctuations, the 
probability density can be rewritten as
\begin{align}
\p=A\exp\left\{-\dfrac{1}{2t}\left[\pfrac{t}{\sn}_{\phi}\(\delta\sn\)^{2}
+\pfrac{\qn}{\phi}_{\sn}\(\delta\phi\)^{2} \right]\right\}.
\label{p2}
\end{align}
Using this form of the probability density, we can get
\begin{align*}
\la\(\delta \phi\)^{2}\ra 
=\dfrac{t}{\pfrac{\qn}{\phi}_{\sn}},\quad 
\la\(\dfrac{\delta \phi}{\phi}\)^{2}\ra 
=\dfrac{\la\(\delta \phi\)^{2}\ra }{\phi^{2}} \propto \dfrac{1}{\n}.
\end{align*}
The mean-square entropy fluctuation remains the same as above, and there is no 
correlation between the entropy and potential fluctuation due to the lack of mixed term 
involving product of $\delta\sn$ and $\delta\phi$ in eq.\eqref{p2}. 

Let us remind that the fluctuation for the local internal energy 
can be obtained by use of the first order relation 
\[
\delta \en=t\delta\sn+\phi\delta\qn.
\]
This can be achieved by first taking the square of the above equation on both sides and 
then calculating the average under the probability distribution $\p$,
\begin{align}
\la(\delta\en)^2\ra = t^2 \la(\delta\sn)^2\ra 
+ 2 t\phi \la\delta\sn \delta\qn\ra 
+ \phi^2 \la(\delta\qn)^2\ra.
\label{de2}
\end{align}

All these procedures seem to be standard as in any text book on thermodynamics. 
However, there is still something to be taken care of. When evaluating the mean-square 
relative fluctuations, we assumed that expressions for
the probability density are of Gaussian type, which means that the exponents 
can be arranged in a quadratic form $- \frac{1}{2t}\sum_{i,j} X_i K_{ij} X_j$, 
where $X_i$ represent the independent 
fluctuations and the matrix $K$ must be positive definite. 
In the first case \eqref{p1}, we have $X=(\delta\sn, \delta\qn)$ and hence
\begin{align}
K=\begin{pmatrix}
\pfrac{t}{\sn}_{\qn}&\pfrac{t}{\qn}_{\sn}\\
\pfrac{t}{\qn}_{\sn}&\pfrac{\phi}{\qn}_{\sn}
\end{pmatrix}. \label{K1}
\end{align}
In the second case, we have $X=(\delta\sn, \delta\phi)$, hence
\begin{align}
K=\begin{pmatrix}
\pfrac{t}{\sn}_{\phi}&0\\
0&\pfrac{\qn}{\phi}_{\sn}
\end{pmatrix}. \label{K2}
\end{align}
The point is that, the positive-definiteness of $K$, or simply ${\rm det}\,K>0$, 
is not guaranteed in either cases. 

Let us look at the first choice \eqref{K1} and take the compact TBHs as our example
system. In this case, we have
\begin{align}
{\rm det}\,K= \pfrac{t}{\sn}_{\qn}\pfrac{\phi}{\qn}_{\sn}-\left[\pfrac{t}{\qn}_{\sn}\right]^2.
\label{detK1}
\end{align}
Notice that ${\rm det}\,K$ as given above coincide precisely with what appears in the 
denominators of eqs.\eqref{fluctq} and \eqref{correlqs}. 
The second term on the RHS of eq.\eqref{detK1} is always non-positive, 
therefore, the positive-definiteness of $K$ will be broken provided 
the first term is negative or vanishes. According to eq.\eqref{esqn}, we have
\begin{align*}
\pfrac{t}{\sn}_{\qn} =\frac{1}{\n} \pfrac{t}{s}_{q},\quad 
\pfrac{\phi}{\qn}_{\sn}= \frac{1}{\n} \pfrac{\phi}{q}_{s}.
\end{align*}
Then it can be inferred from Fig.~\ref{fig4} that $\pfrac{\phi}{q}_{s}$ is always positive, 
while according to the first two columns in Fig.~\ref{fig1}, $\pfrac{t}{s}_{q}$ can be 
negative or zero for spherical BHs with $q<q_c$. When this happens, the positive-definiteness 
of $K$ is broken, and the procedure for calculating the relative mean-square
fluctuations would lead to divergent results. Such strong relative fluctuations 
in the chosen subsystem will destroy the equilibrium state of the whole black hole 
system, until new equilibrium is reached. This process is nothing but the 
phase transition process, as it is evident that the states at which $\pfrac{t}{s}_{q}$ 
takes negative or vanishing values are located exactly in the phase transition zone. 
In fact, the condition for breaking the positive-definiteness of $K$ does not require 
strict non-positivity for $\pfrac{t}{s}_{q}$. For positive, but sufficiently small 
values of $\pfrac{t}{s}_{q}$, ${\rm det}\,K$ can still be non-positive. That is why 
meta-stable states need to be replaced by truly stable states in the coexistence zone
for stable small and large black hole phases.

For the second choice \eqref{K2}, we have
\begin{align}
{\rm det}\,K= \pfrac{t}{\sn}_{\phi}\pfrac{\qn}{\phi}_{\sn}
= \pfrac{t}{s}_{\phi}\pfrac{q}{\phi}_{s}.
\end{align}
Once again, $\pfrac{q}{\phi}_{s}$ is always positive, and thus whether $K$ is 
positive definite depends solely on the sign of $\pfrac{t}{s}_{\phi}$. 
It can be seen in Fig.~\ref{fig3} that, for spherical AdS black holes in Einstein-Maxwell
theory, $\pfrac{t}{s}_{\phi}$ can be negative or zero provided $\phi<1/a$ and 
$0<t\leq t_{\rm min}$, where $t_{\rm min}$ is the lowest temperature of such black holes. 
In such cases, the corresponding relative mean-squared fluctuations also 
diverge, and such divergences signify the non-equilibrium phase transition 
from an unstable small black hole phase to a stable large black hole phase. 
The celebrated HP transition is a special case of these transitions which corresponds 
to a transition not between two black hole branches, but rather between an AdS black hole 
phase and a thermal gas with zero chemical potential.  

It should be mentioned that, when the (relative) fluctuations for the local 
entropy and charge become large, so is $\la(\delta\en)^2\ra$, as is implied by 
eq.\eqref{de2}. This explains why there could be a finite jump in $e$ when first order
phase transition happens.

As an addendum to the above discussions, let us mention that, if the subsystem 
which we choose were open rather than closed, then the Smoluchowski formula
\eqref{prd} would become
\[
\p = A\exp\(-\frac{\delta t\delta\sn+\delta\phi\delta\qn+\delta\bar\mu\delta\n}{2t}\).
\]
However, this form of Smoluchowski formula is barely of any use, 
because, taking any three of the six fluctuation variables and re-express the rest ones 
in terms of them would lead to a distribution of the form
\begin{align}
\p = A\exp\(- \frac{1}{2t} X K X^T\)  \label{Probopen}
\end{align}
with a degenerate matrix $K$. The reason lying in behind is the Gibbs-Duhem relation,
which implies that there are at most two independent fluctuation variables. 

Summarizing the above discussions, let us mention that the local fluctuations 
in small closed subsystems of a black hole can be strong enough to destroy the 
equilibrium state of the entire system, leading to thermodynamic 
instabilities and phase transitions. Looking at the 
places where the coefficient matrix $K$ in the Smoluchowski distribution becomes 
non-positive-definite can be a quick method for finding black hole phase transitions. 
These ideas work only in the presence of Euler homogeneity, which allows for 
the investigation of subsystem behaviors. 

\subsection{The approach with thermodynamic geometry}

There exist different descriptions for thermodynamic geometries in the literature, 
e.g. Weinhold geometry\cite{weinhold1975metric,weinhold2009classical}, 
Ruppeiner geometry\cite{ruppeiner1981application,ruppeiner1995riemannian}, etc. 
The metrics for Weinhold and 
Ruppeiner geometries differ only by a finite Weyl factor, therefore, regarding 
the divergence behaviors, the two descriptions are basically identical. 

Let us take Ruppeiner geometry as an example. The ordinary treatment for 
black hole fluctuations using Ruppeiner geometry starts from defining the metric on the 
space of macro states as
\begin{align}
g_{ab} = -\ppfrac{S}{X^a}{X^b}, \label{Rm}
\end{align}
where $X^a, X^b$ represent the other additive quantities on which the entropy 
$S$ depends. The phase transition points would then be identified as the divergent
points of the Ricci scalar $R(g_{ab})$ associated with the Ruppeiner metric \eqref{Rm}. 
For BHs with finite horizon areas, and when their thermodynamic properties 
are described using the traditional or extended phase space formalisms,  
this approach has been verified for years, which proves to be effective
\cite{aaman2003geometry,aaman2006geometry,
sahay2010thermodynamic,zhang2015phase,
wei2017critical,wei2019ruppeiner,xu2020diagnosis,xu2020ruppeiner,Wc2022}. 
However, for BHs whose thermodynamic behaviors are described using the RPS formalism, 
or for BHs with infinite horizon areas (such as the non-compact TBHs studied 
in this work), there would be some technical problems with the Ruppeiner geometric 
approach. 

The first problem is relatively minor, which is related to the divergence of 
additive quantities for the non-compact TBHs, and can be easily resolved by
considering only finite subsystems. The second problem arises as a consequence of 
the complete Euler homogeneity, or more precisely, the Gibbs-Duhem relation. 
Taking the compact BHs in Einstein-Maxwell theory as an example (which avoid 
the first problem). We can solve the Euler relation to get the entropy as a function in 
$(E,\hat Q, N)$, with the total differential
\begin{align}
\rd S = \frac{1}{T}(\rd E - \hat\Phi\rd\hat Q-\mu\rd N). \label{rdS}
\end{align}
Clearly, we have
\[
\pfrac{S}{E}_{\hat Q,N}=\frac{1}{T},\quad \pfrac{S}{\hat Q}_{E,N}=-\frac{\hat\Phi}{T},
\quad \pfrac{S}{N}_{E,\hat Q}=-\frac{\mu}{T}.
\]
Due to the Gibbs-Duhem relation, only two of the three quantities $(T,\hat\Phi,\mu)$ 
are independent. Therefore, taking $X^a=(E,\hat Q, N)$ would give rise to a degenerate
matrix $g_{ab} = -\ppfrac{S}{X_a}{X_b}$ with ${\rm det}(g_{ab})=0$, which 
cannot be taken as a Riemannian metric. In essence, this problem is of the same nature
as the the problem of degeneracy of the matrix $K$ in eq.\eqref{Probopen}.

There is a single solution to resolve the above two problems altogether, i.e. by
considering only {\em closed subsystems} with fixed $\n$. If we persist in using the 
dimensionless variables, eq.\eqref{rdS} will be replaced with
\begin{align}
\rd\sn=\frac{1}{t}(\rd\en-\phi\rd\qn).
\end{align}
If we take $x^a \equiv (t,\qn)$ as independent variables and consider $\en$ and $\phi$ 
as functions in $(t,\qn)$, the local Ruppeiner-like metric
\[
\gn_{ab}\equiv - \ppfrac{\sn}{x^a}{x^b}
\]
can be evaluated to be
\begin{align}
\gn_{ab}=\frac{1}{t} \begin{pmatrix}
\frac{\cn_\qn}{t} & 0\cr
0& \pfrac{\phi}{\qn}_t
\end{pmatrix}, \label{Rmetric}
\end{align}
where $\cn_\qn=\pfrac{\en}{t}_{\qn}=\n c_q$ is the dimensionless heat capacity of the 
subsystem. The global version of a similar metric was given in \cite{wei2019ruppeiner}, 
which is inapplicable for non-compact TBHs. Here the local version recovers the 
full applicability for the cases of all TBHs with any horizon geometry.

The point lies in that the line element
\begin{align}
\delta^2\sn = - \gn_{ab} \delta x^a \delta x^b
\label{d2s}
\end{align} 
represents the second order fluctuation for the dimensionless entropy $\sn$ 
of the finite subsystem (the first order fluctuation vanishes 
provided the subsystem is in equilibrium with the reservoir). In the stable 
states of the subsystem, $\delta^2\sn$ needs to be negative, which corresponds to
positive values of $\cn_\qn$ and $\pfrac{\phi}{\qn}_t$. However, in the  
unstable states, $\delta^2\sn$ can be non-negative, indicating that the state of the 
subsystem would never fall back to the initial state spontaneously after the 
fluctuation. Moreover, at places where $\cn_\qn$ and $\pfrac{\phi}{\qn}_t$ diverge, 
the local fluctuation becomes infinitely strong, consequently the equilibrium of the 
whole black hole system can be destroyed, leading to a system-wide phase transition. 
Such situations indeed happen, as can be inferred from Figs.~\ref{fig2} and \ref{fig5}.

Now let us make our critical points by asking the following two questions: 

(1) Why Ruppeiner geometry works by identifying phase transition
points with the divergence points of the Ricci scalar associated with the Ruppeiner metric?

(2) \cyan{Do} phase transitions have anything to do with the Riemannian structure on the 
space of macro states? 

The answer to the first question can be given by evaluating the 
Ricci scalar $R(\gn_{ab})$ associated with $\gn_{ab}$ explicitly,
\begin{align}
R(\gn_{ab})
&=\frac{1}{2(\cn_\qn)^2 ({\pt_\qn\phi})^2}\Big\{
t \pt_\qn\phi \big[\pt_t \cn_\qn \big(t \pt_{t,\qn}\phi-\pt_\qn\phi\big)
+(\pt_\qn\cn_\qn)^2\big]\nonumber\\
&\qquad +\cn_\qn \big[
t \big( \pt_\qn \cn_\qn \pt_{\qn,\qn}\phi + t (\pt_{t,\qn}\phi )^2\big)
-2 t \pt_\qn\phi \big(\pt_{\qn,\qn}\cn_\qn + t \pt_{t,t,\qn}\phi\big)
-(\pt_\qn \phi)^2\big]\Big\}.
\label{Rgn}
\end{align}
$R(\gn_{ab})$ could be divergent either at the zeros of the denominator
or at the divergent points of the numerator. 
For $t>0$, the only zero comes from the zero of $\pt_\qn\phi\equiv \pfrac{\phi}{\qn}_t$, 
which corresponds to the existence of horizontal tangent to the isothermal 
$\phi-q$ curves, as can be seen in the first row of Fig.~\ref{fig5}. 
Whether $R(\gn_{ab})$ is indeed divergent at the zero of $\pt_\qn\phi$ 
cannot be determined by looking at eq.\eqref{Rgn} alone without calculating the 
various derivatives appearing in the numerator. However, 
at $t=0$, an extra zero arises in the denominator for some of the TBHs 
with reasonable low temperature limits, because $\cn_\qn$ goes to zero 
either as $\mathcal{O}(t^1)$ or as $\mathcal{O}(t^{d-2})$. In this case, 
$R(\gn_{ab})$ is definitely divergent, because the last term on the second line 
of eq.\eqref{Rgn} gives a contribution $-\frac{1}{2\cn_\qn}$, which diverges at $t=0$.  
This divergence could not be identified as 
a phase transition point, because the low temperature limit of the TBHs --   
when exists -- looks completely normal. This may be a signature for the failure 
of the Ruppeiner geometric approach, as is noticed in \cite{aaman2003geometry, 
aaman2006geometry} in the cases of four dimensional asymptotically flat Kerr black hole
and higher dimensional multiply rotating black holes without 
cosmological constant, for which no phase transitions of any type exist. 
 
The divergent points of the numerator occur at places where $\cn_\qn$ diverges, 
because, at the divergent points of $\cn_\qn$, $\pt_{\qn,\qn} \cn_\qn$ diverges at a
much faster rate than $\cn_\qn$ itself, which leads to a divergence in 
$R(\gn_{ab})$. Using the identity
\begin{align*}
\pt_\qn\phi &=\pfrac{\phi}{\qn}_t 
=\pfrac{\phi}{\qn}_\sn +\pfrac{\phi}{\sn}_\qn \pfrac{\sn}{\qn}_t
=\frac{1}{\n}\left[\pfrac{\phi}{q}_s +\pfrac{\phi}{s}_q \pfrac{s}{q}_t\right]\\
&=\frac{1}{\n}\left[\pfrac{\phi}{q}_{s}
-\pfrac{\phi}{s}_{q}\dfrac{\pfrac{t}{q}_{s}}{\pfrac{t}{s}_{q}}\right]
=\frac{1}{\n}\left[\pfrac{\phi}{q}_{s}
-\frac{c_q}{t}\pfrac{\phi}{s}_{q}\pfrac{t}{q}_{s}\right],
\end{align*}
one can see that $\pt_\qn\phi$ diverges 
precisely at the same states where $\cn_\qn$ diverges. Therefore, there is no need 
to consider the divergence of $R(\gn_{ab})$ caused by the divergence of $\pt_\qn\phi$
in the numerator separately.

In all cases, the divergences of $R(\gn_{ab})$ are actually 
inherited from the divergent or degenerate points of $\gn_{ab}$. 
However, since the Ruppeiner metric is already singular or degenerate at the 
divergent points of $R(\gn_{ab})$, the evaluation of $R(\gn_{ab})$ itself at 
such points is at most formal, and it is unnecessary to do so -- it suffices to 
find the divergent/degenerate points of the Ruppeiner metric, 
which is technically much easier and physically more sounded. Moreover, 
by looking for the places where $\cn_\qn$ and/or $\pfrac{\phi}{\qn}_t$ become 
negative, one can judge the thermodynamic instability of the subsystem. This 
information can be clearly inferred from the Ruppeiner metric but not 
from the corresponding curvature scalar. 

Our answer to the second question listed above tends to be ``No'',  
because the role of general coordinate transformations has never been made 
sufficiently clear in such settings, in spite of the numerous arguments made in 
\cite{ruppeiner1995riemannian}. The point lies in that, each variable in the 
space of macro states has a clear physical meaning, and the singular/degenerate 
points in the Ruppeiner metric also have very clear physical correspondences. 
Therefore, it is not clear what a general coordinate transform on the 
space of macro states means. Especially, when the coordinate transform is nonlinear, 
the Euler homogeneity which is a characteristic property of thermodynamics would be broken. 
The argument \cite{ruppeiner1995riemannian} that the energy and the homogeneity are 
irrelevant to the analysis of fluctuations is incorrect, especially in the 
present local thermodynamic framework, because, in the absence of Euler homogeneity, 
there will be no point to take a subsystem as a representative for investigating   
mean behaviors of the complete system. 
Moreover, the singularities in $\gn_{ab}$ or its inverse 
are different from the coordinate singularities which could appear in spacetime metrics 
in relativistic theories of gravity or in generic Riemannian spaces. 
In the latter cases, the Ricci scalar needs not to inherit 
the singularities in the metric, because some of the singularities may be 
spurious and may be removed by coordinate transformations. 
The use of Riemannian geometric constructions on 
the space of macro states can be confusing and cumbersome, while the use of conventional 
techniques for finding phase transition points is logically clearer and 
technically simpler.

\section{Concluding remarks}

The RPS formalism proves to be a powerful tool for analyzing thermodynamic 
behaviors for TBHs with different horizon geometries. The complete Euler homogeneity
of this formalism is especially useful for understanding the properties of 
non-compact TBHs, which are hard to understand properly using other formalisms 
of black hole thermodynamics. The concept of subsystems for black holes 
introduced in this work allows all conventional approaches in ordinary thermodynamic
systems to be introduced into black hole systems, which in turn helps in analyzing the 
local thermodynamic behaviors of finite parts of the black hole event horizons 
as subsystems. 

Our analysis for the concrete example cases of TBHs in Einstein-Maxwell theory and in 
$4d$ conformal gravity indicate that, the local thermodynamic behaviors for TBHs 
can be quite different for TBHs in the same underlying theory but with different 
horizon geometries. On the other hand, TBHs with the same horizon geometry but
from different underlying gravity models can also behave very differently, 
and there exist special cases in which certain behavior of some black hole solutions 
looks ill-posed from thermodynamic perspective, and this opens a novel possibility 
for detecting the physical viability of gravity models or their solutions 
using thermodynamic considerations. 
 
Besides the detailed comparison for concrete thermodynamic behaviors of 
TBHs with different horizon geometries in the above two models of gravity, 
the most striking new result of this work lies in the high temperature 
and low temperature limits. On the high temperature end, 
our analysis indicates that all TBHs behave like quantum phonon gases 
in nonmetallic crystals residing in a flat space with two less dimensions than the 
black hole spacetime. This behavior is irrespective of the underlying gravity model, 
the spacetime dimension and the concrete horizon geometry. This fact further supports 
our conjecture about the universality of the AdS/phonon gas correspondence. 
On the low temperature end, it is shown that most of the TBHs in $4d$ 
conformal gravity do not admit a low temperature limit, 
except for the neutral planer case which still behaves like $2d$ phonon gases 
as in the high temperature limit;  while in Einstein-Maxwell theory, 
all charged TBHs as well as the neutral hyperbolic TBHs behave like 
strongly degenerated electron gases, and the neutral planar TBHs retain the 
phonon gas behavior. The neutral compact TBHs in Einstein-Maxwell theory do not admit
a low temperature limit. 

The introduction of finite closed subsystems allows for an 
analysis about the local thermodynamic fluctuations, taking the whole black hole with 
(nearly) infinitely many microscopic degrees of freedom as an 
isolated heat reservoir. This view point is very important, because, 
in the absence of the concept of subsystems, the only way to 
consider thermal fluctuations of a black hole might \cyan{be} to immerse it 
in an external heat reservoir. However, such configurations can hardly be trustable, 
the large external heat reservoir will most probably destroy 
the black hole configuration and change it into something else (which may be a 
larger black hole in a different thermodynamic state), rather than 
stay in equilibrium with the black hole. The concrete analysis of the strong 
local mean fluctuations in the presence of instability allows us to understand the 
origin of the finite jump in the mean internal energy per black hole molecule $e$
during the process of first order phase transition, which is otherwise difficult to 
understand in the global picture (in which the jump becomes that of the mass of the 
black hole, which breaks energy conservation). Certain conceptual and technical 
issues in the local version of the Ruppeiner geometric approach to black hole 
fluctuations are also discussed in some detail. 

Last but not the least, let us pay some words on the potential observational 
consequences of the first order black hole phase transitions. Please be aware that,
the first order $t-s$ phase transitions could appear only for BHs 
with a compact event horizon in AdS backgrounds. Since our universe is not asymptotically 
AdS, it appears that there can hardly be any $t-s$ phase transitions for black holes in
the observational universe. However, as pointed out in Refs.\cite{box1,box2,box3}, 
BHs confined in a box can mimic the behaviors of AdS BHs. In reality, a 
confining potential provided by another strong gravitational source can 
play the role of a confining box, thus there is a possibility to observe 
black hole phase transitions in the real universe. Since it has been made clear in the 
main text of this paper, the $t-s$ phase transitions are accompanied by a jump in 
the black hole mass, a black hole in a confining potential will most probably undergo a
transition from a state with larger mass to a state with smaller mass 
during the first order phase transition in the absence of external mass source. 
During this process, some extra mass could be ejected out of the black hole. 
Such mass ejection could in principle be observed and may provide a mechanism 
for explaining the black hole ``burp'' phenomenon reported recently in \cite{burp}.

\section*{Acknowledgement}

This work is supported by the National Natural Science Foundation of China under the grant
No. 12275138.

\providecommand{\href}[2]{#2}\begingroup
\footnotesize\itemsep=0pt
\providecommand{\eprint}[2][]{\href{http://arxiv.org/abs/#2}{arXiv:#2}}


\end{document}